\documentclass[aps,prb,twocolumn,superscriptaddress,floatfix,showpacs]{revtex4}
\usepackage{graphicx}
\usepackage{epsfig}
\usepackage{bm}
\usepackage{times}
\usepackage{amssymb}
\usepackage{url,hyperref}

\begin{document}

\title{The underscreened Kondo lattice model applied to heavy fermion uranium compounds}

\author{N. B. Perkins}
\affiliation{Institute fur Theoretische Physik, TU Braunschweig,
Mendelssohnstrasse 3, 38106 Braunschweig, Germany}
\affiliation{Bogoliubov Laboratory of Theoretical Physics, JINR,
Dubna, Russia}
\author{M. D. N\'{u}\~{n}ez-Regueiro \footnote{Our friend and coworker Mar\'{\i}a Dolores N\'{u}\~{n}ez-Regueiro passed away on November 15, 2006 during the final stages of this work}}
\author{B. Coqblin}
\affiliation{Laboratoire de Physique des Solides, Universit\'{e}
Paris-Sud, UMR-8502 CNRS, 91405 Orsay, France}
\author{J. R. Iglesias}
\affiliation{Instituto de F\'{\i}sica, Universidade Federal do Rio
Grande do Sul, 91501-970 Porto Alegre, Brazil}

\date{\today}

\begin{abstract}
We present theoretical results for the underscreened Kondo lattice
model with localized $S=1$ spins coupled to a conduction band
through a Kondo coupling, $J_K$, and interacting among them
ferromagnetically. We use a fermionic representation for the spin
operators and expand the Hamiltonian in terms of bosonic fields. For
large values of $J_K$, we obtain a ferromagnetically ordered
solution and a Kondo regime with a Kondo temperature, $T_K$, larger
than the Curie temperature, $T_C$. This finding suggests a scenario
for a coexistence of Kondo effect and ferromagnetic order. In some
uranium compounds, like $UTe$ or $UCu_{0.9}Sb_{2}$, this kind of
coexistence has been experimentally observed: they order
ferromagnetically with a Curie temperature of order $T_C \sim 100K$
and exhibit a Kondo behavior for $T > T_C$. The proposed
underscreened Kondo lattice model accounts well for the coexistence
between magnetic order and Kondo behavior and yields to a new
``ferromagnetic Doniach diagram''.
\end{abstract}


\maketitle

\section{\label{sec:intro}INTRODUCTION}

The Kondo lattice (KL) model is one of the fundamental microscopic
models for studying the properties of strongly correlated electron
systems, and a large amount of theoretical work was carried out on
this problem in recent years (for a review, see
Ref.~\onlinecite{Tsurevmod}). This model is widely used to describe
the physics of intermetallic heavy fermion compounds based either on
rare earths elements, or on actinides~\cite{stewart}. In heavy
fermion materials there are two different types of electrons:
conduction electrons from outer atomic orbitals, and strongly
correlated electrons from inner $f$-orbitals, the later ones being
generally localized. The KL model describes the interaction between
these two electronic subsystems in the limit when $f$-electrons are
completely localized and  form a lattice of localized spins.

Historically, KL model has been proposed to account for properties
of cerium compounds, where a competition between Kondo effect and
magnetic order has been experimentally observed. Such  competition
gives rise to a  rich phase diagram with various quantum phase
transitions. KL model has been proven to be an appropriate tool for
describing these quantum transitions at different values of external
parameters such as band filling, pressure, magnetic field or
temperature\cite{CoqblinKLM,CoqblinPhilMag}.

In most cerium compounds, Ce ions are in the localized $4f^1$
configuration corresponding to spin $S=1/2$. This localized spin
couples antiferromagnetically, via an on-site exchange interaction,
$J_K$, to the conduction electron spin density.  At very low
temperatures the localized spin $S=1/2$ is completely screened by
the conduction electrons, leading to the formation of coherent Kondo
spin-singlet state. Besides, the local coupling between $f-$spins
and conduction electrons may give rise to a magnetic order through
the RKKY interaction. This interaction is usually added to the KL
model as an additional inter-site interaction between
$f-$spins~\cite{coleman,Iglesias}.

The competition between the  magnetic order and the Kondo effect was
first considered by Doniach \cite{Doniach,CoqblinKLM}. He proposed a
phase diagram  with a quantum phase transition  between a
magnetically ordered phase and a non-magnetic Kondo phase. Doniach
phase diagram was later extended\cite{CoqblinKLM} to include the
short range magnetic correlations that survive inside the Kondo
phase\cite{Rossat}. From an experimental point of view, the
competition between magnetic order and Kondo effect has been
observed in many cerium and ytterbium compounds, which yields a set
of very rich phase diagrams with various quantum phase transitions
under pressure (see, for example, ref.~\onlinecite{CoqblinPhilMag}
and references therein).

In this paper, we focus on the physical properties of uranium
compounds. This is another class of heavy fermion systems, which
show very rich behavior, quite different from cerium compounds. It
is peculiar of uranium compounds that they exhibit numerous
coexistence phenomena, the most prominent of which is the
coexistence of magnetic order with Kondo
effect~\cite{Schoenes,Schoenes3,Bukowski,Tran,Tran2} or the
coexistence of the magnetic order with
superconductivity~\cite{aoki,flouquet}. We will be primarily
interested  in the coexistence between ferromagnetic order and Kondo
behavior, as, to our knowledge, this effect has been somewhat
overlooked from a theoretical point of view.

Let us briefly describe the experimental situation. The first
experimental evidence of the coexistence between Kondo behavior and
ferromagnetic order in the dense Kondo compound $UTe$ has been
obtained long time ago~\cite{Schoenes}. More recently, this
coexistence has been observed in $UCu_{0.9}Sb_{2}$ ~\cite{Bukowski}
and $UCo_{0.5}Sb_{2}$ ~\cite{Tran,Tran2}. All these systems undergo
a ferromagnetic ordering at the relatively high Curie temperatures
of $T_{C}$ = 102K ($UTe$), $T_{C}$ = 113K ($UCu_{0.9}Sb_{2}$) and
$T_{C}$ = 64.5K ($UCo_{0.5}Sb_{2}$). Above the ordering
temperatures, i.e. in the expected paramagnetic region, these
materials exhibit a Kondo-like logarithmic decrease of the
electrical resistivity, indicating a Kondo behavior. This
logarithmic variation extends down to the ferromagnetic Curie
temperature, $T_C$, suggesting that the Kondo behavior survives
inside the ferromagnetic phase, implying that the ferromagnetic
order and the Kondo behavior do coexist. This coexistence, together
with the large Curie temperatures, are clearly novel features that
cannot be explained by the standard KL model~\footnote{
Nevertheless, we should notice that a kind of coexistence of
ferromagnetism and Kondo-like behavior has been observed too in a
few Ce-based dense Kondo systems, such as CePt$_x$Si compound
\cite{shelton} or CeAg one \cite{Eiling}. But the Curie temperatures
of these compounds are relatively small, typically of order $5K$,
and this result can be considered as a clear sign of a strong
competition rather than a real coexistence between the Kondo effect
and ferromagnetic order}.

Therefore, as a minimal model to describe the Kondo-ferromagnetism
coexistence, we propose an Underscreened Kondo lattice model which,
we argue, is appropriate to describe the $5f^2$ configuration of
uranium ions.

The underscreened Kondo lattice (UKL) model consists of a periodic
lattice of magnetic atoms with $S=1$ interacting with a spin density
of conduction electrons via an on-site antiferromagnetic Kondo
coupling. In addition, the localized spins at neighboring sites
interact ferromagnetically  with each other. In this case the Kondo
effect does not lead to a complete screening of the localized spins,
and the ferromagnetic exchange between the (underscreened) spins may
indeed lead to the formation of ferromagnetic order.

We warn that the choice of the model for the electronic structure of
uranium compounds is a question not settled yet. Magnetism in these
compounds undoubtedly comes from $5f$ electrons -- this has been
proven by many experimental observations, e.g.  by form-factor
studies in neutron scattering. At the same time, $5f$ electron
states in uranium compounds are in a crossover region between
localized and itinerant behavior, and the degree of localization
depends strongly on a subtle balance between the electronic
structure, the effect of correlations and crystal field effects. It
is often difficult to decide, on the basis of the experimental data,
between a local Kondo behavior corresponding to a $5f^n$
configuration and a mixed-valence situation. One example is provided
by uranium monochalcogenides: $US$ lies closest to the itinerant
side for the 5f-electrons, $USe$ is in the middle and $UTe$ is the
closest to the localized side
\cite{ShengCooper,Schoenes2,Schoenes4}. Recent photoemission
experiments on $UTe$ have been interpreted as favoring itinerant
magnetism \cite{Dur}, but the magnetic moments deduced from magnetic
susceptibility experiments in this compound are close to the free
ion values of uranium, which implies that the $5f$ electrons are
relatively well localized in $UTe$ ~\cite{Schoenes2,Schoenes3}.
Moreover, the dual nature of the $5f$ electrons, assuming two
localized 5f electrons and one delocalized one, has been considered
by Zwicknagl et al. \cite{Zwicknagl1,Zwicknagl2}who have obtained by
band calculations a mass enhancement factor in good agreement with
experiment in $UPt_{3}$ and $UPd_{2}Al_{3}$ and by Schoenes et al.
~\cite{Schoenes2,Schoenes4}who have carefully analyzed the variation
of the localization of the 5f-electrons with concentration and
pressure in diluted $US$ and $UTe$. The electronic structure of
uranium and plutonium monochalcogenides has been also studied by
DMFT calculations ~\cite{Licht}. So, the appropriate description of
the electronic structure for uranium compounds is a challenging
problem and depends strongly on the considered system. Here we
restraint ourselves to the study of the UKL model applied to uranium
compounds such as $UTe$, when the uranium ions are relatively well
localized and can be correctly described within a $5f^{2}$
configuration, in which the two $5f$-electrons are bound into spin
$S=1$.

Besides its applicability to  the physics of ferromagnetic uranium
compounds, UKL model is also  interesting on its own. It is one of
theoretical models which can capture the physics of the lattice of
underscreened magnetic moments in a metal, yet it attracted much
less attention than the underscreened Kondo impurity model, for
which there  there exist various theoretical studies\cite{Gan}, and
an exact solution has been obtained by the Bethe ansatz
\cite{Sacra,Schlott}.  There has been only  few studies of the UKL.
The UKL model in the form which we use here was first proposed in
Ref.\onlinecite{perkins06}. Magnetism and superconductivity in one
dimensional UKL model has been studied in Ref.~\onlinecite{natan}.
The pseudogap formation in the UKL model has been studied in a
large-N limit in Ref.\onlinecite{flor}.

The crucial step in our analysis of the UKL model lies in the choice
of the fermionic representation of localized spins $S=1$.  We model
them  by two degenerate $f$-orbitals with one $f$-electron each. At
each site these two electrons  are bound into $S=1$ due to the
strong on-site Hund's coupling.  Our fermionic representation
projects out singlet states and satisfies spin algebra in the
triplet Hilbert subspace\cite{perkins}. In other words, all possible
spin transitions that leave the system  in the triplet subspace,
$|S=1,S_z \rangle$ are equivalently described in terms of auxiliary
fermionic operators.

We show that UKL model exhibits two continuous transitions (more
precisely, sharp crossovers): the first one, at $T=T_K$, to a
non-magnetic Kondo state with a pseudogap in the $f$-electron
density of states, and the second one, at $T=T_C$, to a
ferromagnetic state.  We find that at strong Kondo coupling
ferromagnetism and Kondo effect do coexist. We evaluate the Kondo
screening, magnetic moments of localized spins and conduction
electrons for various band-fillings in a wide range of coupling
parameters and obtain the phase diagram with the regions of
Kondo-ferromagnetic coexistence, non-magnetic Kondo behavior and
pure ferromagnetism. This phase diagram can be considered as a
ferromagnetic "Doniach" diagram for the UKL model.

The paper is organized as follows: In Sec.II, we present the model,
introduce the fermionic representation for spin operators, derive
the Green functions and perform a self-consistent analysis. The
results of the calculations at zero- and finite temperatures are
discussed in Sec.III. In Sec. III.E, we  present the ferromagnetic
"Doniach diagram" for our model. The conclusion contains the
discussion of the main results and the comparison with the
experimental data for the uranium compounds.

\section{Underscreened Kondo lattice model.}

\subsection{Fermionic representation of localized spins.}

We first introduce and discuss in detail the fermionic
representation for localized spins $S=1$, made out of two fermions
on degenerate $f$-orbitals. The two fermions couple into $S=1$ due
to the strong on-site Hund's interaction. We introduce a fermionic
representation in the constrained Hilbert space which contains only
triplet spin states, dropping out all spin singlet states.

The projection of the spin singlet states is justified due to the
correlation nature of $f$-electrons. The strong Hund's coupling
favors the triplet states with energy $E_t$  with respect to the
singlet states with energy $E_s$, and to the states with two
electrons on a single orbital, with even higher energy $E_d$.

Then, considering just states with $S=1$ the transformation between
different $S_z$ ($|1,S_z \rangle$) projections are unambiguously
described in terms of fermionic operators as~\cite{perkins}:
\begin{eqnarray}
\begin{array}{l}
\frac{1}{2}(f_{1\uparrow}^{\dagger}f_{1\uparrow}
f_{2\downarrow}^{\dagger}f_{2\downarrow} +
f_{1\downarrow}^{\dagger}f_{1\downarrow}
f_{2\uparrow}^{\dagger}f_{2\uparrow}+f_{1\uparrow}^{\dagger}f_{1\downarrow}
f_{2\downarrow}^{\dagger}f_{2\uparrow}
+ \\[0.2cm]
f_{1\downarrow}^{\dagger}f_{1\uparrow}
f_{2\uparrow}^{\dagger}f_{2\downarrow})
:|1,0 \rangle\rightarrow |1,0 \rangle \\[0.2cm]
f_{1\uparrow}^{\dagger}f_{1\uparrow}
f_{2\uparrow}^{\dagger}f_{2\uparrow}
: |1,1 \rangle\rightarrow |1,1 \rangle \\[0.2cm]
 f_{1\downarrow}^{\dagger}f_{1\downarrow}
f_{2\downarrow}^{\dagger}f_{2\downarrow}
: |1,-1 \rangle\rightarrow |1,-1 \rangle \\[0.2cm]
\frac{1}{2}(f_{1\uparrow}^{\dagger}f_{1\uparrow}
f_{2\downarrow}^{\dagger}f_{2\uparrow} +
f_{1\downarrow}^{\dagger}f_{1\uparrow}
f_{2\uparrow}^{\dagger}f_{2\uparrow})
 : |1,1 \rangle\rightarrow |1,0 \rangle \\[0.2cm]
\frac{1}{2}(f_{1\downarrow}^{\dagger}f_{1\downarrow}
f_{2\downarrow}^{\dagger}f_{2\uparrow}+
f_{1\downarrow}^{\dagger}f_{1\uparrow}
f_{2\downarrow}^{\dagger}f_{2\downarrow})
 : |1,0 \rangle\rightarrow |1,-1 \rangle \\[0.2cm]
\frac{1}{2}(f_{1\downarrow}^{\dagger}f_{1\downarrow}
f_{2\uparrow}^{\dagger}f_{2\downarrow}+
f_{1\uparrow}^{\dagger}f_{1\downarrow}
f_{2\downarrow}^{\dagger}f_{2\downarrow})
 : |1,-1 \rangle\rightarrow |1,0 \rangle \\[0.2cm]
\frac{1}{2}(f_{1\uparrow}^{\dagger}f_{1\uparrow}
f_{2\uparrow}^{\dagger}f_{2\downarrow} +
f_{1\uparrow}^{\dagger}f_{1\downarrow}
f_{2\uparrow}^{\dagger}f_{2\uparrow})
 : |1,0 \rangle\rightarrow |1,1 \rangle .
\end{array}
\label{s}
\end{eqnarray}
\noindent where $f_{\sigma\alpha}^{\dagger}$ and $f_{\sigma\alpha}$
are creation and annihilation operators for $f$-electrons, carrying
spin and orbital indexes,  $\sigma$ and  $\alpha(\alpha=1,2)$,
respectively. In terms of spin operators, the transitions
represented by Eqs.(\ref{s}) can be expressed as:
\begin{eqnarray}
\begin{array}{l}
1-S_z^2 : |1,0 \rangle\rightarrow |1,0 \rangle \\[0.2cm]
\frac{1}{2}S_z(S_z-1) : |1,-1 \rangle\rightarrow |1,-1 \rangle \\[0.2cm]
\frac{1}{2}S_z(S_z+1) : |1,1 \rangle\rightarrow |1,1 \rangle
\\[0.2cm]
-\frac{1}{2}S_zS^- : |1,0 \rangle\rightarrow |1,-1 \rangle \\[0.2cm]
-\frac{1}{2}S^{+}S_z : |1,-1 \rangle\rightarrow |1,0 \rangle \\[0.2cm]
\frac{1}{2}S^-S_z : |1,1 \rangle\rightarrow |1,0 \rangle \\[0.2cm]
\frac{1}{2}S_zS^{+} : |1,0 \rangle\rightarrow |1,1 \rangle
\end{array}
\label{s1}
\end{eqnarray}
Note that there is no transition $|1,1\rangle\leftrightarrow |1,-1
\rangle$. The equivalence between fermionic and spin representations
should be interpreted as the equivalence of matrix elements between
corresponding states.

\subsection{Model.}

Our model Hamiltonian is the following:
\begin{eqnarray}\label{Ham}
H = \sum_{{\bf k}\sigma} (\epsilon_{\bf k}-\mu)c^{\dagger}_{{\bf
k}\sigma}c_{{\bf k}\sigma}+
\sum_{i\sigma\alpha}E_{o}n_{i\sigma}^{f_{\alpha}}+\\\nonumber J_K
\sum_{i}{\bf S}_i{\bf{\sigma}}_i + \frac{1}{2}J_H \sum_{ij}{\bf
S}_i{\bf S}_j
\end{eqnarray}
\noindent The first term represents the conduction band with
dispersion energy $\epsilon_{k}$. We assume that the band has a
width $2D$ and the  density of states for conduction electrons is
$1/2D$ in the interval $[-D,D]$ and zero otherwise. The operators
$c_{{\bf k}\sigma}^{\dagger} (c_{{\bf k}\sigma})$ correspond to
delocalized Bloch states with spin $\sigma$, while $\mu$ is the bare
electron chemical potential. The second term describes the energy of
localized levels and $E_o$ can be considered as a fictitious
chemical potential, i.e. a Lagrange multiplier for auxiliary
$f$-fermions. The actual value of $E_o$ is fixed by a local
constraint $n_{f}=\sum_{i\sigma\alpha}n_{i\sigma}^{f_{\alpha}}=2$
for the number of $f$-electrons per site. The third  term is the
antiferromagnetic on-site Kondo coupling, with $J_{K}>0$, between
localized $f$-spins, $S_i=1$, and conduction electrons
$\sigma_i=1/2$ spins. The spin operators $\sigma_i$ can be written
in terms of the fermionic operators in a standard way: $
\sigma_i^+=c_{i\uparrow}^{+}c_{i\downarrow},~~
\sigma_i^-=c_{i\downarrow}^{+}c_{i\uparrow},~~
\sigma_i^z=\frac{1}{2}(n^c_{i\uparrow}-n^c_{i\downarrow})$. The last
term in Eq.(\ref{Ham}) is a ferromagnetic inter-site interaction,
$J_{H}<0$, between localized $f$-magnetic moments~\cite{f}.

Using the fermionic representation  for both $f$- and conduction
electrons spins, we can rewrite the Hamiltonian (\ref{Ham}) in terms
of $f$ and $c$ electronic operators. As the expression of the
Hamiltonian in terms of fermionic operators is rather lengthy, we
show here in detail just the most relevant term, the transverse part
of the Kondo coupling
$\frac{1}{2}J_K(\sigma_i^+S_i^-+\sigma_i^-S_i^+)$, which reads, when
expressed in the fermionic representation:
\begin{eqnarray}
\begin{array}{l}
\frac{1}{2}J_K{\Big (}\sigma_i^+S_i^-+\sigma_i^-S_i^+{\Big )}=
\frac{1}{2}J_K{\Big (}c_{i\uparrow}^{+}c_{i\downarrow}(
f_{i1\uparrow}^{+}f_{i1\uparrow} f_{i2\downarrow}^{+}f_{i2\uparrow}
+ \\[0.2cm]
f_{i1\downarrow}^{+}f_{i1\uparrow} f_{i2\uparrow}^{+}f_{i2\uparrow}
+ f_{i1\downarrow}^{+}f_{i1\downarrow}
f_{i2\downarrow}^{+}f_{i2\uparrow}+
f_{i1\downarrow}^{+}f_{i1\uparrow}
f_{i2\downarrow}^{+}f_{i2\downarrow})
+\\[0.2cm]
c_{i\downarrow}^{+}c_{i\uparrow}(
f_{i1\downarrow}^{+}f_{i1\downarrow}
f_{i2\uparrow}^{+}f_{i2\downarrow}+
f_{i1\uparrow}^{+}f_{i1\downarrow}
f_{i2\downarrow}^{+}f_{i2\downarrow}+\\[0.2cm]
f_{i1\uparrow}^{+}f_{i1\uparrow} f_{i2\uparrow}^{+}f_{i2\downarrow}
+ f_{i1\uparrow}^{+}f_{i1\downarrow}
f_{i2\uparrow}^{+}f_{i2\uparrow} ){\Big )} \label{kt1}
\end{array}
\end{eqnarray}
Now we define the relevant bosonic fields to describe the Kondo
effect: we introduce the operator
$\widehat{\lambda}_{i\sigma}=\sum_{\alpha}
\widehat{\lambda}_{i\sigma}^{\alpha}=\sum_{\alpha}
c_{i\sigma}^{+}f_{i\sigma}^{\alpha}$ which couples electrons and
auxiliary fermions  at the same site.  Using this definition, we
rewrite the Eq.(\ref{kt1}) as:
\begin{eqnarray}
\begin{array}{l}
\frac{1}{2}J_K{\Big (}\sigma_i^+S_i^-+\sigma_i^-S_i^+{\Big )}=
\frac{1}{2}J_K{\Big (}
-\lambda_{i\uparrow}^{2}\lambda_{i\downarrow}^{*2}n_1-
\lambda_{i\uparrow}^{1}\lambda_{i\downarrow}^{*1}n_2
-\\[0.2cm]
\lambda_{i\downarrow}^{2}\lambda_{i\uparrow}^{*2}n_1-
\lambda_{i\downarrow}^{1}\lambda_{i\uparrow}^{*1}n_2 {\Big )}=
-\frac{1}{2}J_K\sum_{\alpha\sigma}
\lambda_{i\sigma}^{\alpha}\lambda_{i\bar{\sigma}}^{*{\alpha}}
n_{\bar{\alpha}} \label{kt2}
\end{array}
\end{eqnarray}

And in order to describe the magnetic properties of the system, we
introduce the operators of magnetization for both $f$- and $c$-
subsystems:
$M_i=S_i^z=\frac{1}{2}(n_{i\uparrow}^f-n^f_{i\downarrow})$ and
$m_i=\sigma_i^z=\frac{1}{2}(n^c_{i\uparrow}-n^c_{i\downarrow})$,
respectively.

We restrict our consideration to self-consistent analysis
(equivalent to  slave boson mean-field description), and evaluate
all  physical quantities in terms of these bosonic fields. We then
introduce four real order parameters
$\lambda_{\sigma}=\langle\widehat{\lambda}_{i\sigma}\rangle$, $M
=\langle {M_i}\rangle$ and $m=\langle {m_i}\rangle$, where
$\langle...\rangle$ denotes the thermal average. The non-zero values
of  $\langle M\rangle$ and $\langle m\rangle$ describe the magnetic
phase with non-zero total magnetization, while a non-zero
$\lambda_\sigma$ describes the Kondo effect and the formation of the
heavy-fermion state. The limits of this approach are discussed at
the end of this subsection.

Within this mean field (MF) approximation, and using the fact that
the total number of f-electrons in each sublevel is $n_{\alpha}=1$,
the Hamiltonian (\ref{Ham}) is expressed in terms of the four order
parameters $\langle\lambda_\sigma\rangle$, $\langle M\rangle$ and
$\langle m\rangle$ as
\begin{eqnarray}
\begin{array}{l}
H= \sum_{{\bf k} \sigma} \varepsilon_{{\bf k}\sigma} c_{{\bf k}
\sigma}^+c_{{\bf k}\sigma} +
 \sum_{i\alpha\sigma}
E_{0\sigma }n_{i\alpha\sigma }-\\[0.2cm]
-\frac{1}{2}J_K \sum_{i\alpha\sigma}
(\langle\lambda_{\bar{\sigma}}\rangle \lambda_{i{\sigma}}^{{\alpha}}
+ h.c.)
+\\[0.2cm]
2J_K N \sum_{\sigma} \langle\lambda_{\bar{\sigma}}\rangle
\langle\lambda_{{\sigma}}\rangle - J_K N\langle m \rangle\langle M
\rangle -\frac{1}{2}J_H N z\langle M \rangle ^2
\end{array}
\label{Ham1}
\end{eqnarray}
\noindent where
\begin{eqnarray}
\begin{array}{l}
\varepsilon_{{\bf k}\sigma} =\varepsilon_{\bf k}+\Delta_{\sigma}~,
~~E_{0\sigma } =E_0+\Lambda_{\sigma}~,\\[0.2cm]
\Delta_{\sigma}=J_K\sigma\langle M \rangle~, ~~\sigma=\pm 1/2\\
\Lambda_{\sigma}= J_K\sigma\langle m \rangle-\frac{1}{2}J_K
\langle\lambda_{{\sigma}}\rangle
\langle\lambda_{\bar{\sigma}}\rangle +J_H z \sigma\langle M \rangle
\end{array}
\label{def1}
\end{eqnarray}
\noindent being $z$ the number of nearest neighbors. In the presence
of non-zero magnetization the bands for up- and down- spins in both
$f$- and $c$- subsystems are shifted from each other: parameters
$\Delta_{\sigma}$ and $\Lambda_{\sigma}$ define the energy shift for
itinerant bands and localized $f$-levels, respectively. The shifts
for $f$- and $c$- bands are in opposite directions because of the
antiferromagnetic Kondo coupling between them.

\begin{figure}[t]
\centerline{\includegraphics[width=10.cm, clip=true]{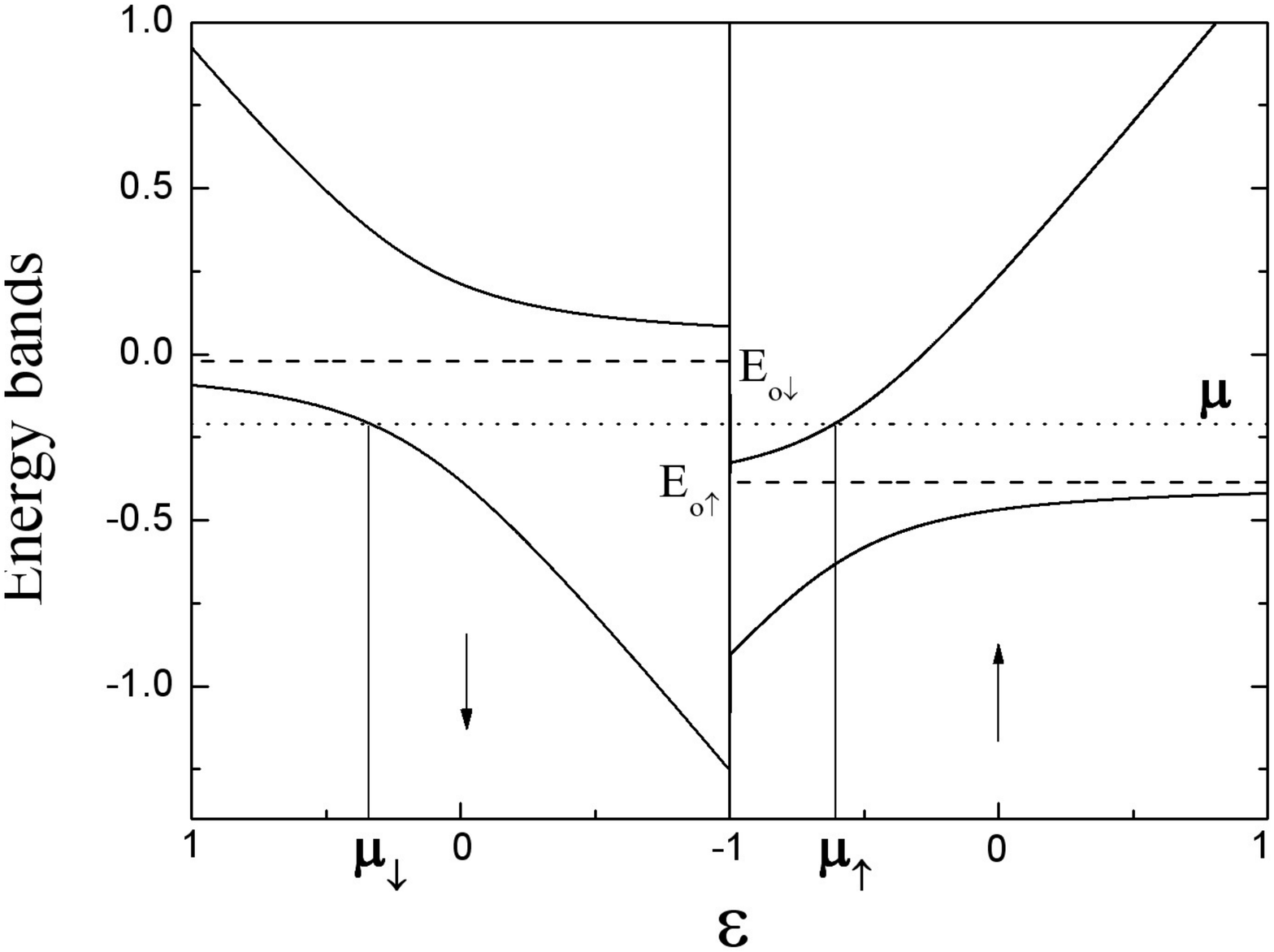}}
\caption{Schematic plot of the band structure: for each direction of
spin there are two  hybridized $E^{\sigma}_{+}$ and
$E^{\sigma}_{-}(k)$  bands (thick solid lines) and one
 non-hybridized $f$-level $E_{0\sigma}$ (dashed line).
The left panel corresponds to down-spins and the right one to
up-spins. The thin dotted line corresponds to the Fermi level.}
\label{bands}
\end{figure}

For each direction of spin, the diagonalization of the Hamiltonian
(\ref{Ham1}) yields one non-hybridized $f$-state with energy
$E_{0\sigma }$ and two hybridized bands with the  quasiparticle
energy dispersion given by:
\begin{eqnarray}
E^{\sigma}_{\pm}({\bf k})=\frac{1}{2}[ E_{0\sigma}+\varepsilon_{{\bf
k}\sigma}\pm \sqrt{(E_{0\sigma}-\varepsilon_{{\bf k}\sigma})^2+8
\alpha^2_{\bar{\sigma}}  }] \label{spectrum}
\end{eqnarray}
\noindent where $\alpha_{\bar{\sigma}}=
-\frac{1}{2}J_K\langle\lambda_{\bar{\sigma}}\rangle$. The $\pm$ sign
refers to the upper and lower hybridized band. The hybridization gap
 in the energy spectrum
$\Gamma_{\sigma}=2\sqrt{2} \alpha
_{\bar{\sigma}}=\sqrt{2}J_K\lambda_{\bar{\sigma}}$ is spin-dependent
and is only present as long as $\lambda_{\sigma}$ is nonzero. The
schematic plot of the band structure is presented in
Fig.\ref{bands}.

The energy spectra $E^{\sigma}_{\pm}({\bf k})$ depend on a set of
external parameters such as the Kondo coupling, $J_K$, and the
exchange interaction, $J_H$, the conduction band filling $n_c$, the
temperature $T$, and a set of internal parameters, $\mu$ and $E_0$,
which should be calculated self-consistently.

The present approach is equivalent to other MF theories developed
for the $S=1/2$  KL model, such as  saddle-point approximation in
the path-integral calculation, performed by Coleman and
Andrei\cite{coleman}, or large-{\it N}, slave boson saddle-point
approximation by Burdin et al~\cite{burdin02}. All these approximate
methods require some caution. Particularly, MF approximation
neglects magnetic fluctuations, and the system may possess spurious
charge fluctuations such that the exact constraint
$f^{\dagger}_{\uparrow}f_{\uparrow}+f^{\dagger}_{\downarrow}f_{\downarrow}=2$
is actually satisfied only on average.

The validity of MF and slave boson approaches for KL model with
localized spin $S=1/2$ has been extensively discussed in
Refs.~\onlinecite{burdin02,burdin01}. It has been shown that the MF
procedure becomes exact when the original model is extended to
$N\rightarrow\infty$ different flavors of localized electrons.
Still, it was found in the same paper that the MF procedure captures
the correct low-temperature physics  even for the actual case of
$N=2$ ($S=1/2)$). By the same reasons, the MF approximation should
work in our case where $N=3$ ($S=1$).

Below we will study the model (\ref{Ham1}) by employing equations of
motion for the Green function (GF) for localized and itinerant
spins. All propagators will be written in the fermionic
representation.

\subsection{Green functions}

We introduce the following notation for the retarded Green
functions: $F_{\alpha\alpha}^{\sigma}= \langle\langle
f_{\alpha\sigma};f_{\alpha\sigma}^{\dagger} \rangle\rangle$,
$F_{\alpha\beta}^{\sigma}= \langle\langle
f_{\alpha\sigma};f_{\beta\sigma}^{\dagger} \rangle\rangle$ for
localized $f$-electrons ; $G_{cc}^{\sigma}= \langle\langle c_{{\bf
k}\sigma};c_{{\bf k}\sigma}^{\dagger} \rangle\rangle$ for itinerant
$c$-electrons, and $G_{c\alpha}^{\sigma}= \langle\langle
c_{\sigma};f_{\alpha\sigma}^{\dagger} \rangle\rangle$ and $G_{\alpha
c}^{\sigma}= \langle\langle f_{\alpha\sigma};c_{\sigma}^{\dagger}
\rangle\rangle$ for the "mixed" $cf$- states. As it is easier to
solve the self-consistent system of equations in the momentum space,
we evaluated all Green functions in ${\bf k}$-representation. The
equations of motion for these four Green functions  are given by
\begin{eqnarray}
\begin{array}{l}
(\omega-E_{o\sigma})F_{\alpha\alpha}^{\sigma}({\bf k})=
1+\alpha_{\bar{\sigma}}G_{c\alpha}^{\sigma}({\bf k})
\\[0.2cm]
(\omega-E_{o\sigma})F_{\beta\alpha}^{\sigma}({\bf k})=
\alpha_{\bar{\sigma}}G_{c\alpha}^{\sigma}({\bf k})
\\[0.2cm]
(\omega-\varepsilon_{{\bf k}\sigma} )G_{c\alpha}^{\sigma}({\bf k})=
\alpha_{\bar{\sigma}}(F_{\alpha\alpha}^{\sigma}({\bf
k})+F_{\beta\alpha}^{\sigma}({\bf k}))
\\[0.2cm]
(\omega-\varepsilon_{{\bf k}\sigma} )G_{cc}^{\sigma}({\bf k})=
1+\alpha_{\bar{\sigma}}(G_{\alpha c}^{\sigma}({\bf k})+G_{\beta
c}^{\sigma}({\bf k}))
\end{array}
\label{gf1}
\end{eqnarray}
\noindent From equations of motion (Eqs.(\ref{gf1})),  by
straightforward but nevertheless tedious algebra (which we report in
the Appendix) we obtain the following expressions for the Green
functions:
\begin{eqnarray}
\begin{array}{l}
F_{\alpha\alpha}^{\sigma}({\bf k})= \frac{1}
{2(\omega-E_{o\sigma})}-\frac{1} {2W_{\sigma}({\bf k})} {\Big [}
\frac{\varepsilon_{{\bf k}\sigma}-E^{\sigma}_{+}({\bf k})}
{\omega-E^{\sigma}_{+}({\bf k})}- \frac{\varepsilon_{{\bf
k}\sigma}-E^{\sigma}_{-}({\bf k})}
{\omega-E^{\sigma}_{-}({\bf k})}{\Big ]}\\[0.3cm]
F_{\beta\alpha}^{\sigma}({\bf k})= -\frac{1}
{2(\omega-E_{o\sigma})}-\frac{1} {2W_{\sigma}({\bf k})} {\Big [}
\frac{\varepsilon_{{\bf k}\sigma}-E^{\sigma}_{+}({\bf k})}
{\omega-E^{\sigma}_{+}({\bf k})}- \frac{\varepsilon_{{\bf
k}\sigma}-E^{\sigma}_{-}({\bf k})} {\omega-E^{\sigma}_{-}({\bf
k})}{\Big ]}
\\[0.3cm]
G_{c\alpha}^{\sigma}({\bf k})= \frac{\alpha_{\bar\sigma}}
{W_{\sigma}({\bf k})} {\Big [} \frac{1} {\omega-E^{\sigma}_{+}({\bf
k})}- \frac{1} {\omega-E^{\sigma}_{-}({\bf k})}{\Big ]}
\\[0.3cm]
G_{cc}^{\sigma}({\bf k})= -\frac{1} {W_{\sigma}({\bf k})} {\Big [}
\frac{E_{o\sigma}-E^{\sigma}_{+}({\bf k})}
{\omega-E^{\sigma}_{+}({\bf k})}-
\frac{E_{o\sigma}-E^{\sigma}_{-}({\bf k})}
{\omega-E^{\sigma}_{-}({\bf k})}{\Big ]}
\end{array}
\label{gf11}
\end{eqnarray}
\noindent where $E^{\sigma}_{\pm}({\bf k})$ are given by
(\ref{spectrum}) and
\begin{eqnarray}
W_{\sigma}({\bf k})\equiv E^{\sigma}_{+}({\bf
k})-E^{\sigma}_{-}({\bf k}) =\sqrt{(E_{o\sigma}-\varepsilon_{{\bf
k}\sigma})^2+8 \alpha^2_{\bar{\sigma}}}~. \label{W}
\end{eqnarray}

\noindent The Green functions given by Eqs.(\ref{gf11}) will be
evaluated self-consistently in the next subsection.

\subsection{Self-consistent equations}

We next construct a close self-consistent scheme to evaluate the
bosonic fields $\lambda_{\uparrow}$, $\lambda_{\downarrow}$, $M$ and
$m$ together with the chemical potential $\mu$ and the Lagrange
multiplier $E_o$. This can be done by imposing (i) constrains on the
total number of $f$-electrons, $n_f=2$, and $c$-electrons, $n_c$,
correspondingly, (ii) the relation between the Fermi surface volume
and the number of particles, (iii) the relation between the total
magnetization and the number of electrons, and (iv) self-consistent
equations for $\lambda_{\uparrow}$ and $\lambda_{\downarrow}$.

The numbers $n_f$ and $n_c$ are expressed via the integrals of the
imaginary parts of the corresponding Green's functions:
\begin{eqnarray}
\begin{array}{l}
n_{f}^{\sigma}= \int_{E_{min}}^{\mu}d \omega
\sum_{{\bf k}\alpha}[-\frac{1}{\pi}Im F_{\alpha\alpha}^{\sigma}({\bf k})]\\[0.2cm]
n_{c}^{\sigma}= \int_{E_{min}}^{\mu}d \omega \sum_{\bf k} {\Big
[}-\frac{1}{\pi}Im G_{cc}^{\sigma}({\bf k}){\Big ]}
\end{array}
\label{s7}
\end{eqnarray}
\noindent where the summation is over all $k$ points of the
Brillouin zone and the orbital indexes $\alpha=1,2$. As the
$k$-dependence in Eqs.(\ref{spectrum}) and (\ref{s7}) comes only
through the bare conduction-band energies $\varepsilon_{{\bf k}
\sigma}$, we can substitute the summation in the $k$-space by the
integration over $\varepsilon_{\sigma }$. The integration is over
the interval $[-D+\Delta_{\sigma}, D+\Delta_{\sigma}]$ where the
density of states $\rho_0=\frac{1}{2D}$ is a constant.

We first consider the case $T=0$ when the calculations can be done
analytically. Let's first evaluate the number of $f$-electrons. For
each polarization of spin, $f$-electrons can occupy one localized
level $E_{o\sigma}$ and a fraction of the two hybridized bands
$E^{\sigma}_{\pm}({\bf k})$. The expression for spin-up
$n_{f}^{\uparrow}$ can be then written as
\begin{eqnarray}
\begin{array}{l}
n_{f}^{\uparrow}=1+ \frac{1}{2D}\int_{-D +\Delta_{\uparrow}}^{D
+\Delta_{\uparrow}} \frac{\varepsilon_{\uparrow}
-E^{\uparrow}_{-}(\varepsilon_{\uparrow})}
{W_{\uparrow}(\varepsilon_{\uparrow})}
 d\varepsilon_{\uparrow}
-\\[0.2cm]
\frac{1}{2D}
\int_{-D+\Delta_{\uparrow}}^{\mu_{\uparrow}+\Delta_{\uparrow}}
\frac{\varepsilon_{\uparrow}
-E^{\uparrow}_{+}(\varepsilon_{\uparrow})}
{W_{\uparrow}(\varepsilon_{\uparrow})} d\varepsilon_{\uparrow} =
\\[0.2cm]
\frac{7}{4}+\frac{\mu_{\uparrow}}{4D}+
\frac{1}{4D}\int_{\mu_{\uparrow}+\Delta_{\uparrow}}^{D+\Delta_{\uparrow}}
\frac{\varepsilon_{\uparrow} -E_{0\uparrow}}
{W_{\uparrow}(\varepsilon_{\uparrow})} d\varepsilon_{\uparrow}~.
\end{array}
\label{s2}
\end{eqnarray}
\noindent All parameters in (\ref{s2}) were defined in (\ref{def1})
and (\ref{W}). We have also  used the following relations:
\begin{eqnarray}
\begin{array}{l}
\varepsilon_{\uparrow}
-E^{\uparrow}_{-}(\varepsilon_{\uparrow})\equiv
\frac{1}{2}(\varepsilon_{\uparrow}-E_{0\uparrow}+W_{\uparrow}(\varepsilon_{\uparrow}))\\[0.2cm]
\varepsilon_{\uparrow}
-E^{\uparrow}_{+}(\varepsilon_{\uparrow})\equiv
\frac{1}{2}(\varepsilon_{\uparrow}-E_{0\uparrow}-W_{\uparrow}
(\varepsilon_{\uparrow}))~.
\end{array}
\label{s3}
\end{eqnarray}
Taking into account that $\int_{a}^{b}\frac{x
dx}{\sqrt{x^2+c}}=\sqrt{x^2+c}~{\Big |}_{a}^{b}$, we obtain:
\begin{eqnarray}
\begin{array}{l}
n_{f}^{\uparrow}= \frac{7}{4}+\frac{\mu_{\uparrow}}{4D}+\frac{1}{4D}
{\Big [}\sqrt{(D
+\Delta_{\uparrow}-E_{0\uparrow})^2+8\alpha_{\downarrow}^2}
\\[0.2cm]-\sqrt{( \mu_{\uparrow}+\Delta_{\uparrow}-E_{0\uparrow})^2+8\alpha_{\downarrow}^2}
{\Big ]}
\end{array}
\label{s5}
\end{eqnarray}
The quantity $\mu_{\uparrow}$, present in Eqs.(\ref{s2}-\ref{s5}) as
the limit of integration over $\epsilon_{\uparrow}$, is related to
$\mu$ through the relations $E^{\uparrow}_{+}(\mu_{\uparrow})=\mu$.

By the same procedure we obtain the expressions for
$n_{f}^{\downarrow}$, $n_{c}^{\uparrow}$ and $n_{c}^{\downarrow}$:
\begin{eqnarray}
\begin{array}{l}
n_{f}^{\downarrow}= \frac{1}{4}+\frac{\mu_{\downarrow}}{4D}-
\frac{1}{4D} {\Big [}\sqrt{(-D+\Delta_{\downarrow}
-E_{0\downarrow})^2+8\alpha_{\uparrow}^2}
\\[0.2cm]-\sqrt{( \mu_{\downarrow}+\Delta_{\downarrow}-E_{0\downarrow})^2+
8\alpha_{\uparrow}^2} {\Big ]}~;\\[0.3cm]
n_{c}^{\uparrow}= \frac{3}{4}+\frac{\mu_{\uparrow}}{4D}-
\frac{1}{4D}{\Big [}
\sqrt{(D+\Delta_{\uparrow}-E_{0\uparrow})^2+8\alpha_{\downarrow}^2}\\[0.2cm]
-\sqrt{(\mu_{\uparrow}+\Delta_{\uparrow}-E_{0\uparrow})^2+8\alpha_{\downarrow}^2}{\Big
]}~;\\[0.3cm]
n_{c}^{\downarrow}= \frac{1}{4}+\frac{\mu_{\downarrow}}{4D}
-\frac{1}{4D}{\Big [}
\sqrt{(\mu_{\downarrow}+\Delta_{\downarrow}-E_{0\downarrow})^2+
8\alpha_{\uparrow}^2}
\\[0.2cm]-
\sqrt{(-D+\Delta_{\downarrow}-E_{0\downarrow})^2+8\alpha_{\uparrow}^2}{\Big
]}~.
\end{array}
\label{s10}
\end{eqnarray}
where $\mu_{\downarrow}$ is obtained from
 $E^{\downarrow}_{-}(\mu_{\downarrow})=\mu$.

Now we can construct the system of self-consistent equations. The
first one is the constraint on the total number of $f$ electrons:
\begin{eqnarray}
n_{f} = n_f^{\uparrow}+ n_f^{\downarrow} =2.
 \label{nf2}
\end{eqnarray}
\noindent As we already discussed above, we replaced the local
constraint $n_{fi}=2$ at each site $i$ by a softer one, for the
average $n_f$.

The second  self-consistent equation is obtained by setting the
average number of conduction electrons  to be equal to the filling
value $n_c$:
\begin{eqnarray}
n_{c}=n_{c}^{\uparrow}+n_{c}^{\downarrow}~. \label{nc2}
\end{eqnarray}

The third and forth self-consistent equations are obtained from the
Luttinger theorem, and from the condition that the total
magnetization $M_{tot}$ is the sum of average magnetization of
$f$-and $c$-electrons. This gives
\begin{eqnarray}
n_{f}+n_{c}=3+ \frac{\mu_{\uparrow}+\mu_{\downarrow}}{2D} ~.
\label{nc22}
\end{eqnarray}
and
\begin{eqnarray}
M+m=1+\frac{\mu_{\uparrow}-\mu_{\downarrow}}{4D}~. \label{nc23}
\end{eqnarray}
We used here $M = \frac{1}{2}(n_{f}^{\uparrow} -
n_{f}^{\downarrow})$ and
$m=\frac{1}{2}(n_{c}^{\uparrow}-n_{c}^{\downarrow})$, respectively.
We remark that the Fermi surface encloses both conduction electrons
and partially localized $f$-levels.

The last two equations are self-consistent relations for the bosonic
fields $\lambda_{\uparrow}, \lambda_{\downarrow}$. These two fields
are related to the imaginary part of the mixed $cf$-Green function:
\begin{eqnarray}
\begin{array}{l}
\lambda_{\sigma}= \int_{E_{min}}^{\mu}d \omega \sum_{\bf
k}[-\frac{1}{\pi}(Im G_{c1}^{\sigma}({\bf k})+Im
G_{c2}^{\sigma}({\bf k}))]~,
\end{array}
\label{l2}
\end{eqnarray}
Using (\ref{gf11}) we then obtain the following expressions for
$\lambda_{\sigma}$:
\begin{eqnarray}
\begin{array}{l}
\begin{array}{c}
\lambda_{\uparrow}= -2\frac{\alpha_{\downarrow}}{2D} {\Big [
}\int_{-D+\Delta_{\uparrow}}^{D+\Delta_{\uparrow}}
\frac{d\varepsilon_{\uparrow}}{W_{\uparrow}(\varepsilon_{\uparrow})}-
\int_{-D+\Delta_{\uparrow}}^{\mu_{\uparrow}+\Delta_{\uparrow}}
\frac{d\varepsilon_{\uparrow}}{W_{\uparrow}(\varepsilon_{\uparrow})}]
\\[0.2cm]
=-\frac{\alpha_{\downarrow}}{D}
\int_{\mu_{\uparrow}+\Delta_{\uparrow}}^{D+\Delta_{\uparrow}}
\frac{d\varepsilon_{\uparrow}}
{W_{\uparrow}(\varepsilon_{\uparrow})}
\end{array}
\\[0.3cm]
\begin{array}{c}
\lambda_{\downarrow}= -\frac{\alpha_{\downarrow}}{D}
\int_{-D+\Delta_{\downarrow}}^{\mu_{\downarrow}+\Delta_{\downarrow}}
\frac{d\varepsilon_{\downarrow}}
{W_{\downarrow}(\varepsilon_{\downarrow})}
\end{array}
\end{array}
\label{l4a}
\end{eqnarray}
Using  $\int \frac{dx}{\sqrt{x^2+c}}= \ln(x+\sqrt{x^2+c})$ we find
\begin{eqnarray}
\begin{array}{l}
\lambda_{\uparrow}=-\frac{\alpha_{\downarrow}}{D}
\ln\frac{D+\Delta_{\uparrow}-E_{0\uparrow}+W_{\uparrow}
(D+\Delta_{\uparrow})}
{\mu_{\uparrow}+\Delta_{\uparrow}-E_{0\uparrow}+W_{\uparrow}
(\mu_{\uparrow}+\Delta_{\uparrow})}
\\[0.2cm]
\lambda_{\downarrow}=-\frac{\alpha_{\uparrow}}{D}
\ln\frac{\mu_{\downarrow}+\Delta_{\downarrow}-E_{0\downarrow}
+W_{\downarrow}(\mu_{\downarrow}+\Delta_{\downarrow})}
{-D+\Delta_{\downarrow}-E_{0\downarrow}+W_{\downarrow}(-D+\Delta_{\downarrow})}
\end{array}
\label{l6}
\end{eqnarray}

Equations (\ref{nf2}),(\ref{nc2}),(\ref{nc22}),(\ref{nc23}) and
(\ref{l6}) constitute the full set of self-consistent equations for
six variables: $\lambda_{\uparrow},~ \lambda_{\downarrow}, ~M, ~m,
~\mu$ and $E_0$.  We solved this equations numerically by iteration
and we explicitly verified that the solution minimizes the total
internal energy $E_{tot}$
\begin{eqnarray}
\begin{array}{l}
E_{tot}=\frac{7}{4}E_{0\uparrow}+ \frac{1}{4D}{\Big [}
\mu_{\uparrow}E_{0\uparrow}+
(\mu_{\downarrow}+D)E_{0\downarrow}-\\[0.2cm]
D^2+2D(\Delta_{\uparrow}-\Delta_{\downarrow})+
\frac{\mu_{\uparrow}^2+\mu_{\downarrow}^2}{2}+
\Delta_{\uparrow}\mu_{\uparrow}+\Delta_{\downarrow}\mu_{\downarrow}
{\Big ]}
\\[0.2cm]
-\frac{1}{8D}{\Big [}
(D+\Delta_{\uparrow}-E_{0\uparrow})W_{\uparrow}(D+\Delta_{\uparrow})-\\[0.2cm]
(\mu_{\uparrow}+\Delta_{\uparrow}-E_{0\uparrow})W_{\uparrow}
(\mu_{\uparrow}+\Delta_{\uparrow})\\[0.2cm]
+
(D+\Delta_{\uparrow}+E_{0\downarrow})W_{\downarrow}(-D+\Delta_{\downarrow})+\\[0.2cm]
(\mu_{\downarrow}+\Delta_{\downarrow}-E_{0\downarrow})W_{\downarrow}
(\mu_{\downarrow}+\Delta_{\downarrow}){\Big ]}
+J_K\lambda_{\downarrow}\lambda_{\uparrow}\\[0.2cm]
-J_K m M -\frac{1}{2}J_Hz M^2 -E_F n_c-2E_0
\end{array}
\label{e11}
\end{eqnarray}

At finite temperature, $T$, the set of self-consistent equations
remains the same, however we are no longer able to obtain an
analytical expression for the occupation numbers and bosonic fields.
The relations between $n_f^\sigma$, $n_c^\sigma$, $\lambda_\sigma$
and the Green's functions are now:
\begin{eqnarray}
\begin{array}{l}
n_{f}^{\sigma}= \int_{-\infty}^{+\infty}d \omega n_F(\omega)
\sum_{\bf k}[-\frac{1}{\pi}Im (F_{11}^{\sigma}({\bf k})+F_{22}^{\sigma}({\bf k}))]\\[0.2cm]
n_{c}^{\sigma}= \int_{-\infty}^{+\infty}d \omega n_F(\omega)
\sum_{\bf k} [-\frac{1}{\pi}Im G_{cc}^{\sigma}({\bf k})]
\\[0.2cm]
\lambda_{\sigma}= \int_{-\infty}^{+\infty} d \omega n_F(\omega)
\sum_{\bf k}[-\frac{1}{\pi}(Im G_{c1}^{\sigma}({\bf k})+Im
G_{c2}^{\sigma}({\bf k}))],
\end{array}
\label{s13}
\end{eqnarray}
\noindent where $n_F(\omega)=\frac{1}{e^{\frac{\omega -\mu}{T}}+1}$
is the Fermi distribution function. Straightforward calculations
lead to
\begin{eqnarray}
\begin{array}{l}
n_{f}^{\sigma}= \frac{1}{2D} \int_{-D +\Delta_{\sigma}}^{D
+\Delta_{\sigma}} d\varepsilon_{\sigma} [ n_F(E_{o\sigma})-
n_F(E_{+}^{\sigma}(\varepsilon_{\sigma})) \frac{\varepsilon_{\sigma}
- E_{+}^{\sigma}(\varepsilon_{\sigma})}
{W_{\sigma}(\varepsilon_{\sigma})} \\[0.2cm]+
n_F(E_{-}^{\sigma}(\varepsilon_{\sigma}))
\frac{\varepsilon_{\sigma}-E_{-}^{\sigma}(\varepsilon_{\sigma})}
{W_{\sigma}(\varepsilon_{\sigma})} ]
\\[0.4cm]
n_{c}^{\sigma}= \frac{1}{2D} \int_{-D +\Delta_{\sigma}}^{D
+\Delta_{\sigma}}d\varepsilon_{\sigma} [
-n_F(E_{+}^{\sigma}(\varepsilon_{\sigma}))
\frac{E_{o\sigma}-E_{+}^{\sigma}(\varepsilon_{\sigma})}
{W_{\sigma}(\varepsilon_{\sigma})}\\[0.2cm]+
n_F(E^{\sigma}_{-}(\varepsilon_{\sigma}))
\frac{E_{o\sigma}-E_{-}^{\sigma}(\varepsilon_{\sigma})}
{W_{\sigma}(\varepsilon_{\sigma})}]
\\[0.4cm]
\lambda_{\sigma}= \frac{1}{D}\int_{-D +\Delta_{\sigma}}^{D
+\Delta_{\sigma}}d\varepsilon_{\sigma}
[n_F(E_{+}^{\sigma}(\varepsilon_{\sigma}))-
n_F(E_{-}^{\sigma}(\varepsilon_{\sigma}))]
\frac{\alpha_{\bar\sigma}}{W_{\sigma}(\varepsilon_{\sigma})}
\end{array}
\label{lambdasigma}
\end{eqnarray}

As for $T=0$, we now evaluate numerically the integrals in
Eqs.(~\ref{lambdasigma}) and verified that the numerical solution of
self-consistent equations minimizes the free energy of the system
$F$. The latter can be calculated through the partition function
$Z$:
\begin{eqnarray}
Z & = & \sum_{{\bf k}\sigma}[e^{-\beta (E^{\sigma}_{-}({\bf
k})-\mu)}+e^{-\beta (E^{\sigma}_{+}({\bf k})-\mu)}]\\ \nonumber & &
+ \sum_{\sigma}e^{-\beta (E_{o\sigma}-\mu)}
\label{partition}
\end{eqnarray}
and
\begin{eqnarray}
\begin{array}{l}
F=-T\ln Z
\end{array}
\label{freeen}
\end{eqnarray}

In the following section, we present the detailed discussion of the
results obtained with this self-consistent scheme for the  UKL
model.

\section{Results.}

\subsection{T=0}
\begin{figure}[t]
\centerline{\includegraphics[width=9.cm,
clip=true]{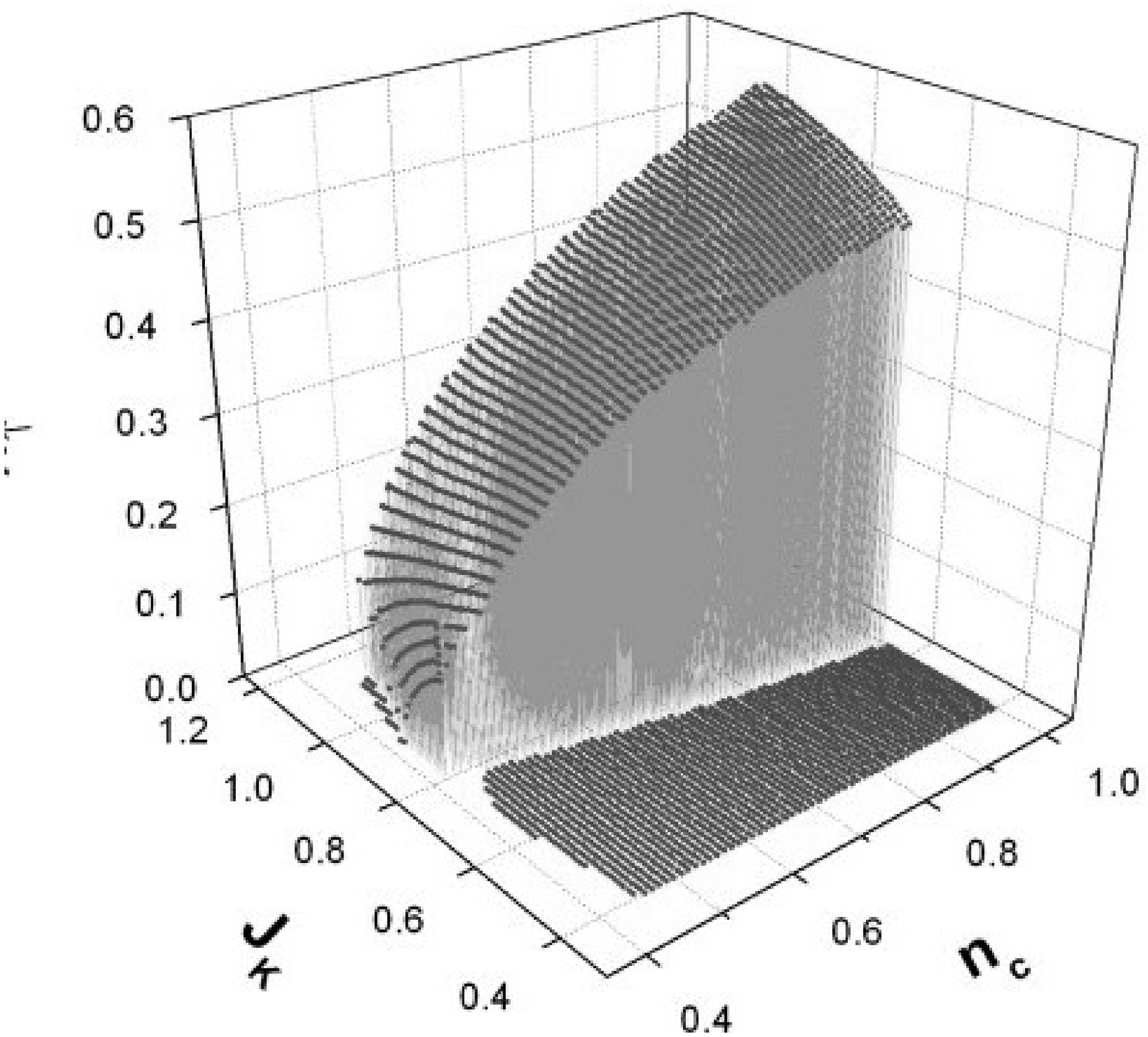}} \caption{Plot of the parameter
$\lambda_{\uparrow}$ as a function of $J_K$ and $n_c$ for
$J_H=-0.01$. There is a discontinuous transition as a function of
$J_K$ and a behavior in $n_c^{1/2}$ as explained in the text}
\label{lambda1}
\end{figure}

\begin{figure}[t]
\centerline{\includegraphics[width=9.cm, clip=true]{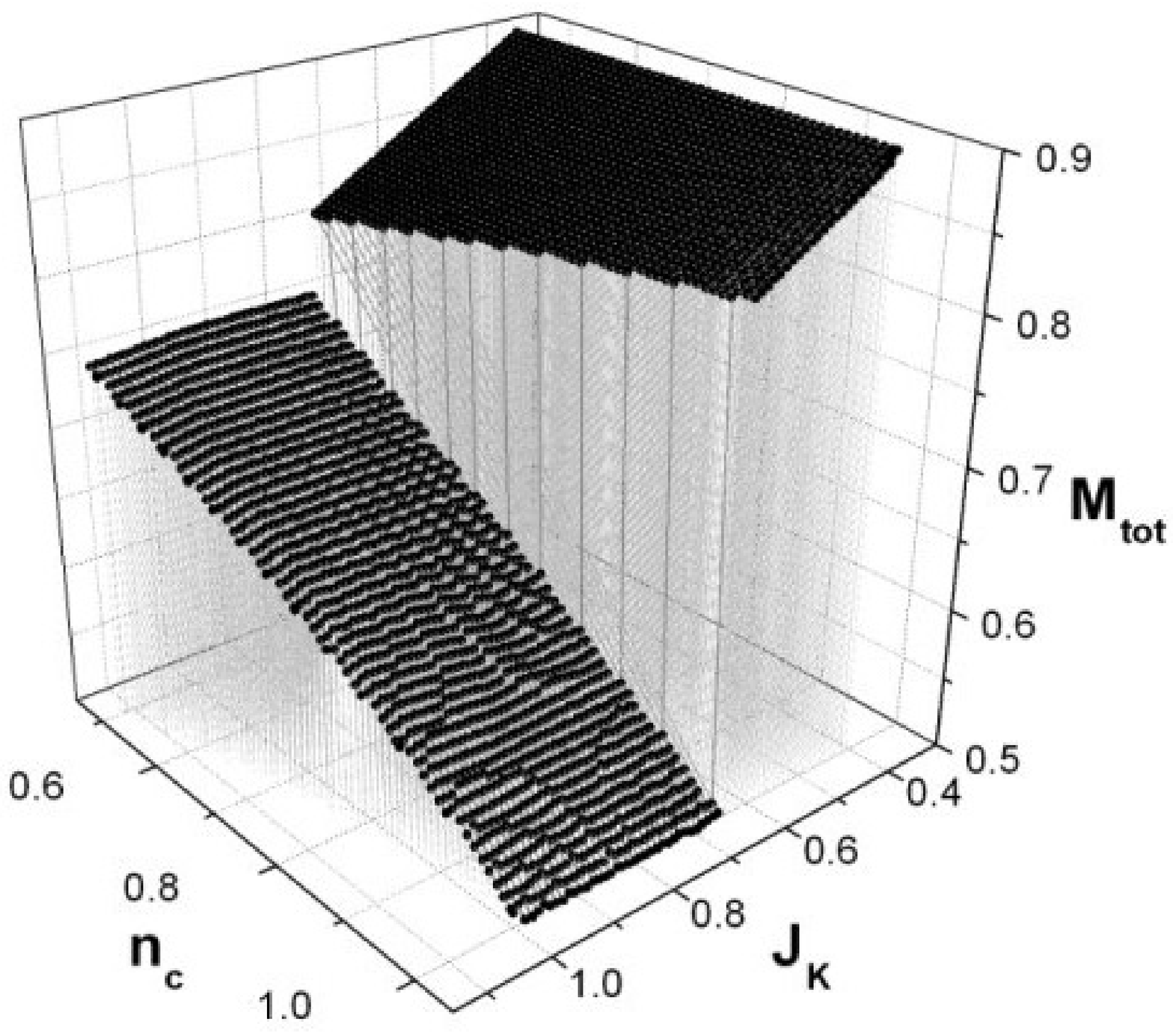}}
\caption{Plot of the total magnetization ($M+m$) as a function of
$J_K$ and $n_c$ for $J_H=-0.01$. There is a discontinuous transition
as a function of $J_K$ that corresponds to the transition to
$\lambda_{\uparrow}=0$ in Fig.\ref{lambda1}}. \label{Mtot}
\end{figure}

We first discuss the properties of the model at $T=0$. In order to
establish  the region of coexistence of Kondo effect and
ferromagnetic ordering (Kondo-ferromagnetism coexistence) in the
phase space set by the parameters of the model, we study the
behavior of  $\lambda_{\sigma}$ and the magnetization $M$ and $m$.
These order parameters are correlated and it is useful to discuss
them together. In Fig.\ref{lambda1} we present the variation of the
Kondo correlation $\lambda_{\uparrow}$ as a function of the Kondo
coupling, $J_{K}$, and the band-filling, $n_{c}$, for a fixed value
of $J_H=-0.01$. In Fig.\ref{Mtot} we present the dependence of the
total magnetization ($M+m$) on the same parameters. In these and all
other figures all energies and temperatures are measured in units of
the half-bandwidth $D$. The values of $\lambda_{\downarrow}$ and
$\lambda_\uparrow$ are close, therefore we present just
$\lambda_\uparrow$ in Fig.~\ref{lambda1} in order to keep the figure
transparent.

From the analysis of both figures one can verify that for values of
$J_{K}$ smaller than 0.6, the Kondo correlation $\lambda{\uparrow}$
remains equal to zero (Fig.\ref{lambda1}) for all values of $n_{c}$
considered. The total magnetization $M+m$ is  large and equal to its
maximum value (Fig.\ref{Mtot}). This phase is a pure magnetic one
with no Kondo effect. When $J_{K}$ increases, for a fixed value of
$n_{c}$, there exists a critical value of $J_K$ above which both
$\lambda_\uparrow$ and the total magnetization $M_{tot}$ change
abruptly and $\lambda_\uparrow$ become different from zero. The
region of the parameters $J_K$ and $n_c$ where $\lambda_\uparrow$
and the total magnetization $M_{tot}$ are finite is the region of
coexistence between Kondo effect and ferromagnetic order. As the
magnetization of localized and conduction electrons, $M$ and $m$,
have opposite signs due to antiferromagnetic Kondo coupling between
them, the maximum value of the total magnetization is always less
than $1$. The critical value of $J_K$  decreases with increasing
$n_{c}$, as it is expected from the "exhaustion principle"
\cite{Nozieres,CoqblinKLM,burdin02}.

\begin{figure}[t]
\centerline{\includegraphics[width=9.cm, clip=true]{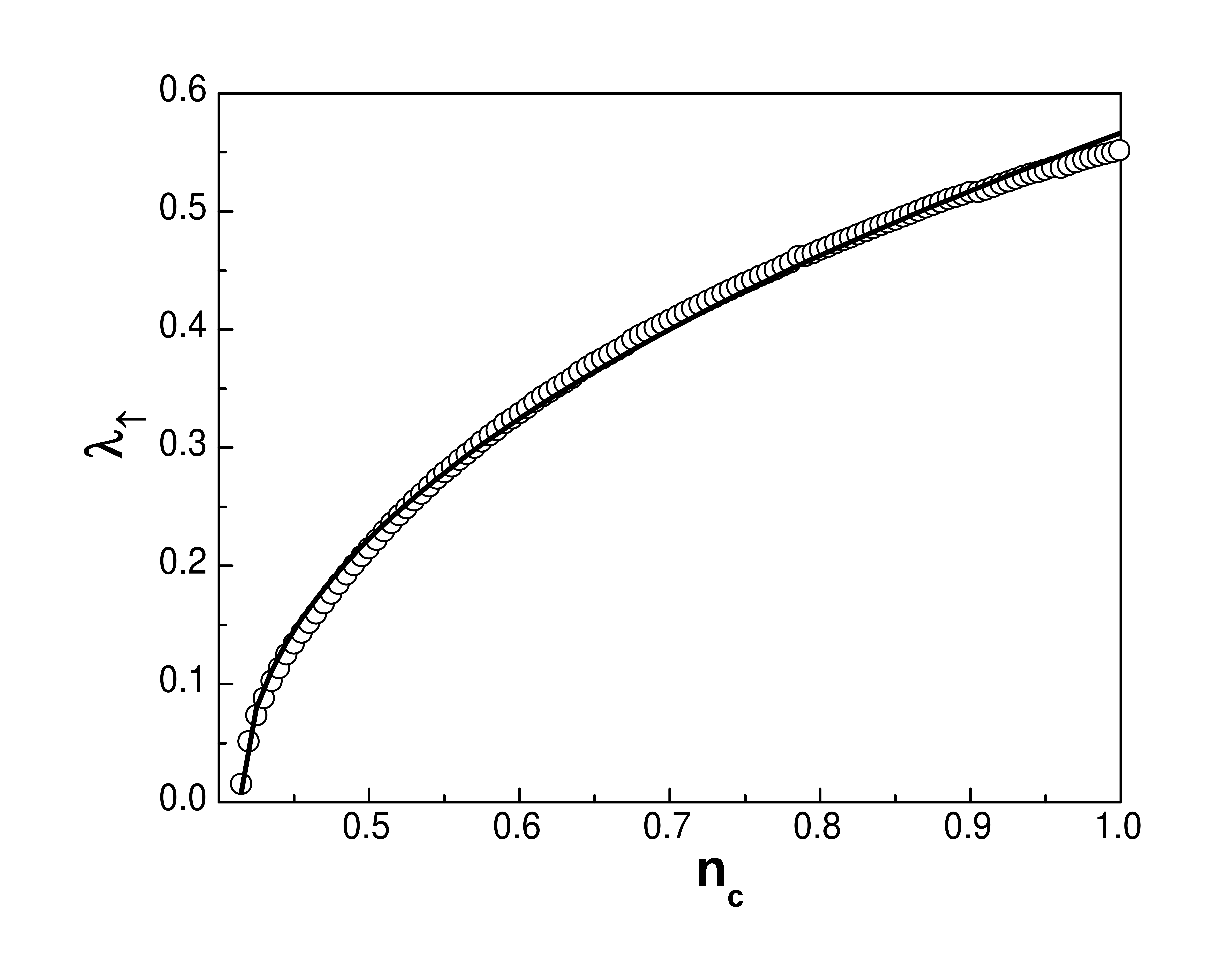}}
\caption{Plot of the parameter $\lambda_{\uparrow}$ as a function of
$n_c$, for $J_K=0.9$ and $J_H=-0.01$. The circles correspond to the
numerical calculation and the line to a fit with a square root law.
$\lambda_{\uparrow}$ goes to zero for $n_c \backsimeq 0.42$
}\label{bestfit}
\end{figure}

One can see in Fig.\ref{lambda1} that $\lambda_\uparrow$ decreases
smoothly as a function of $n_c$, while it undergoes a sharp
transition as a function of $J_K$. We calculated the variation of
$\lambda$ with $n_c$ and  obtained an approximate square root
behavior $\lambda \sim (n_c - n_{c, cr})^r$, with $r \approx 1/2$,
similar to the KL case with spin $1/2$~\cite{CoqblinKLM,burdin02}.
Indeed, in Fig. \ref{bestfit} we present a plot of
$\lambda_{\uparrow}$ as a function of $n_c$ for $J_K=0.9$ and
$J_H=-0.01$. The solid line in this figure is the best fit to the
numerical results and yields a power law behavior with an exponent
very near $1/2$.

\begin{figure}[t]
\centerline{\includegraphics[width=9.cm]{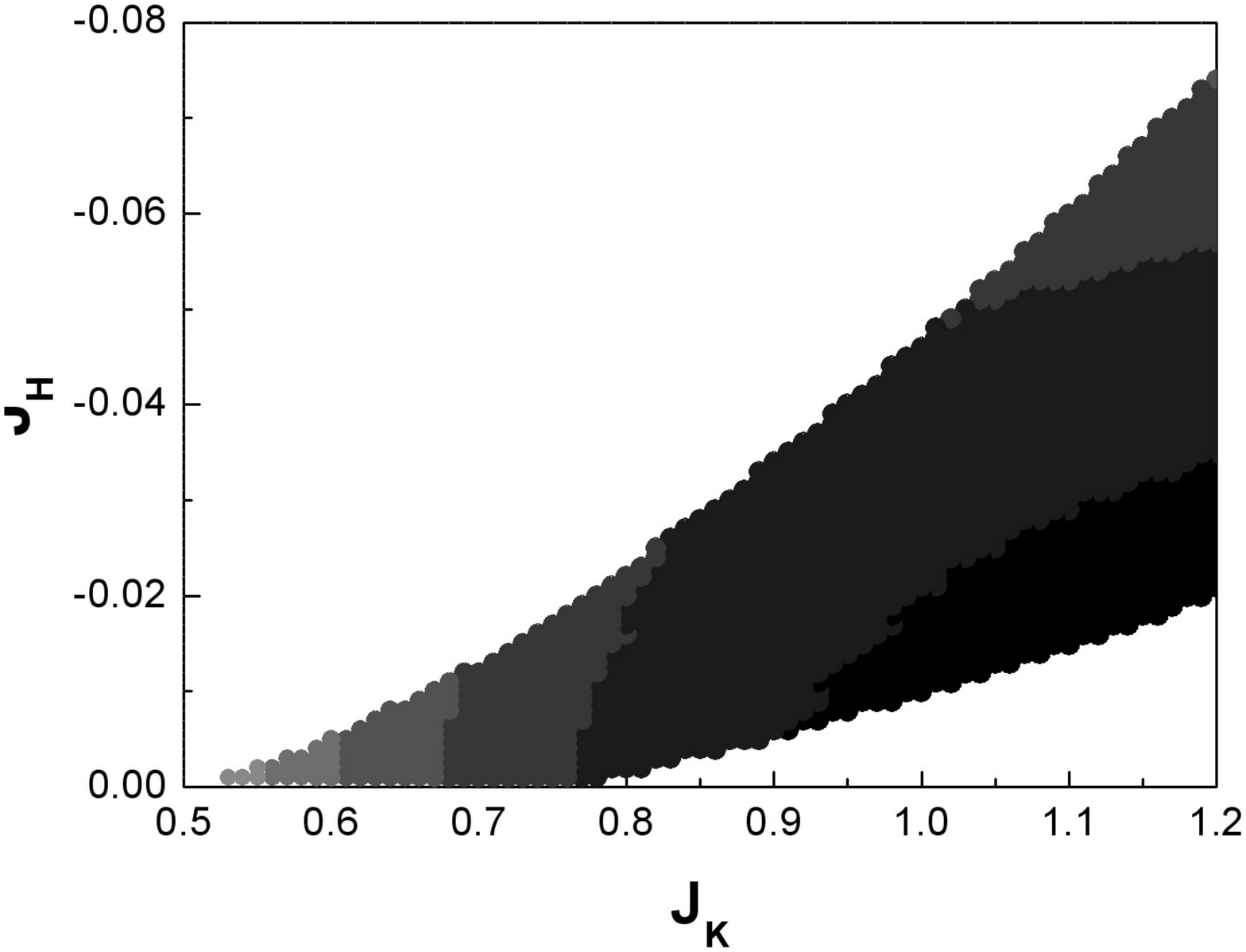}}
\caption{The dark region indicates the values of the parameters
$J_K$ and $J_H$ where there is coexistence between the magnetic
ordered state and the strongly correlated (Kondo) state. The
intensity of the gray tone is proportional to the value of
$\lambda_{\uparrow}$} \label{phasediag}
\end{figure}

We next study the effect of the exchange interaction between
localized spins, $J_H$, on the region of the Kondo-ferromagnetism
coexistence. In Fig.\ref{phasediag} we present the phase diagram in
the $\{J_{K},J_{H}\}$ plane for a fixed value of $n_{c}$. The region
of coexistence extends up to  $J_{K} =1.2$ and down to $J_{K} =0.5$
while $J_{H}$ can vary between $-0.001$ and $-0.06$. The values of
$J_H$ in Fig.\ref{phasediag} are rather small; however we recall
that the real strength of the ``local field'' applied on an
individual spin by its neighbors is $zJ_{H}$, where $z=6$ is the
number of nearest neighbors in a simple cubic lattice.

Summarizing, at $T=0$ there exists a region of coexistence between
the Kondo effect and the ferromagnetic order in a wide range of
parameters $J_K$, $J_H$, and $n_{c}$, including the case of the
half-filled conduction band, $n_c=1$. Also, a discontinuous
transition to pure ferromagnetic state is found by decreasing $J_K$
while a continuous change is obtained varying the band filling,
$n_c$.

\subsection{Finite temperatures}

\begin{figure}[t]
\centerline{\includegraphics[width=9.cm, clip=true]{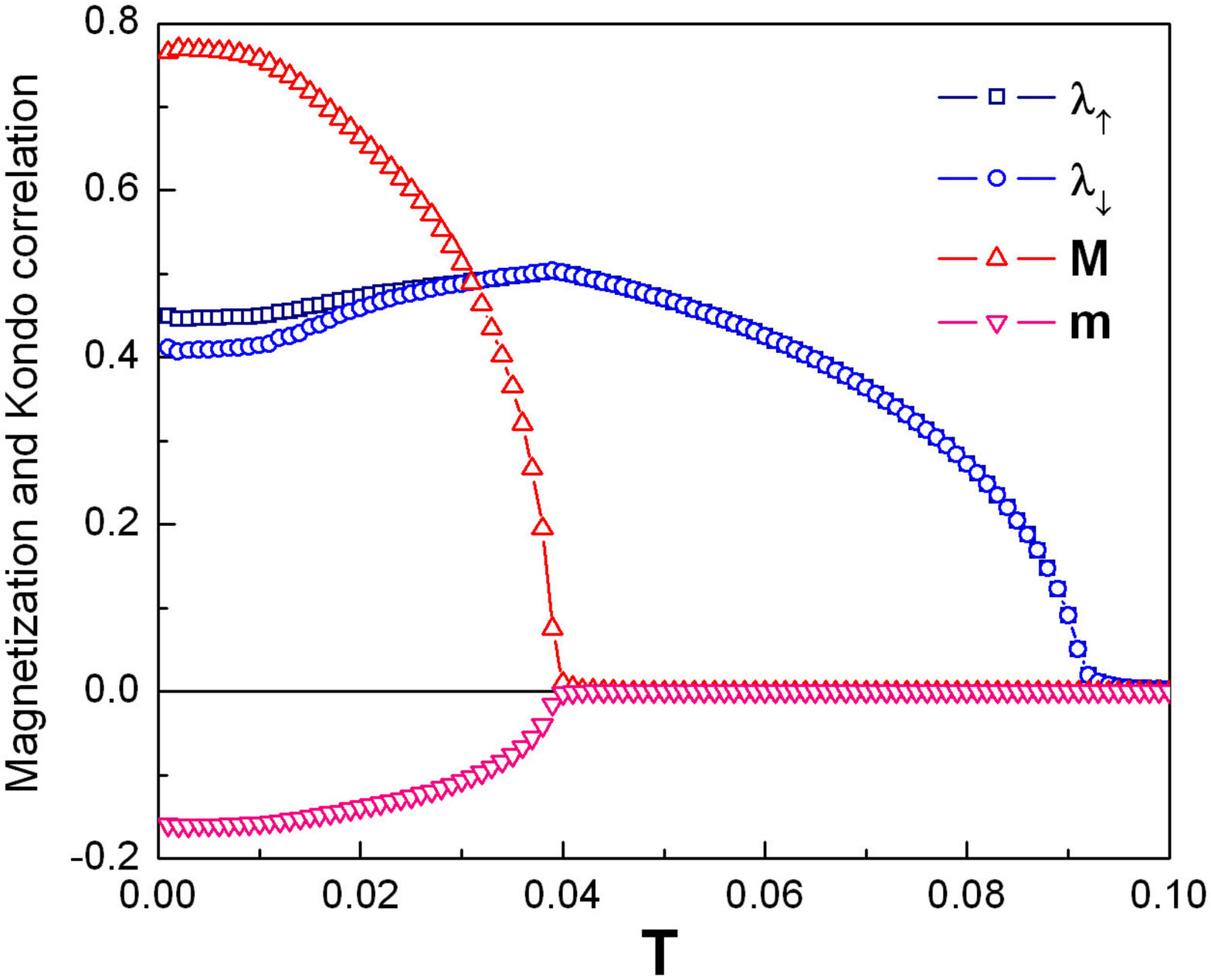}}
\caption{Plot of $\lambda_{\uparrow}$, $\lambda_{\downarrow}$, $M$
and $m$ as a function of temperature for $J_K=0.8$, $J_H=-0.01$,
$n_c=0.8$. At zero and low temperatures we observe the coexistence
of magnetic order and heavy fermion behavior, the Kondo effect is
maximal when magnetization vanishes.} \label{temp}
\end{figure}

We now present the  results obtained at finite temperatures. First
we show in Fig.~\ref{temp} the temperature variation of the Kondo
correlations $\lambda_{\uparrow}$ and $\lambda_{\downarrow}$, and
also the $f$ and $c$ magnetization $M$ and $m$. The parameters used
in the calculation of Fig.~\ref{temp} are: $J_{K} = 0.8$, $J_{H} = -
0.01$ and $n_{c} = 0.8$. The two magnetization curves clearly show
the existence of a continuous phase transition at the Curie, $T_C$,
from a ferromagnetic to a paramagnetic state. At low temperatures we
observe the coexistence of a magnetic order and Kondo behavior. The
strength of the Kondo effect is the highest at $T_C$. As a matter of
fact, when the magnetization decreases, $\lambda_{\sigma}$ grows and
passes through a maximum at the $T_C$; its value at that temperature
is about $20\%$ bigger than at $T=0$. Due to the breakdown of the
spin symmetry, $\lambda_{\downarrow}$ and $\lambda_{\uparrow}$ are
slightly different in the magnetic region but they coincide at
$T_C$, when the spin symmetry is restored. For $T>T_C$ the system
exhibits only Kondo behavior ($\lambda_\sigma \neq 0$, $M=0$ and
$m=0$). Finally, we define the temperature at which $\lambda_\sigma$
vanishes and $f$ and $c$ electrons become decoupled, as the Kondo
temperature, $T_K$. The fact that $\lambda$ vanishes at a particular
temperature, instead of slowly decaying to zero, is a well known
artifact of the mean-field approximation. Actually, $T_K$ is a
crossover temperature, associated with the onset of local Kondo
screening.

\subsection{Density of states}

We also analyze the $f$- and $c$- electron densities of states (DOS)
at various temperatures. The $f$- and $c$- electron DOS are
calculated numerically from the imaginary part of the $f-f$ and
$c-c$ Green functions.
\begin{eqnarray}
\rho_f(\omega) & = & \sum_{k\alpha}[-\frac{1}{\pi}Im
F_{\alpha\alpha}^{\sigma}] \label{fdos1}, \\
\rho_c(\omega)& = & \sum_{k}[-\frac{1}{\pi}Im G_{cc}^{\sigma}]
\label{fdos2}
\end{eqnarray}

\begin{figure}[t]
\centerline{\includegraphics[width=9.cm, clip=true]{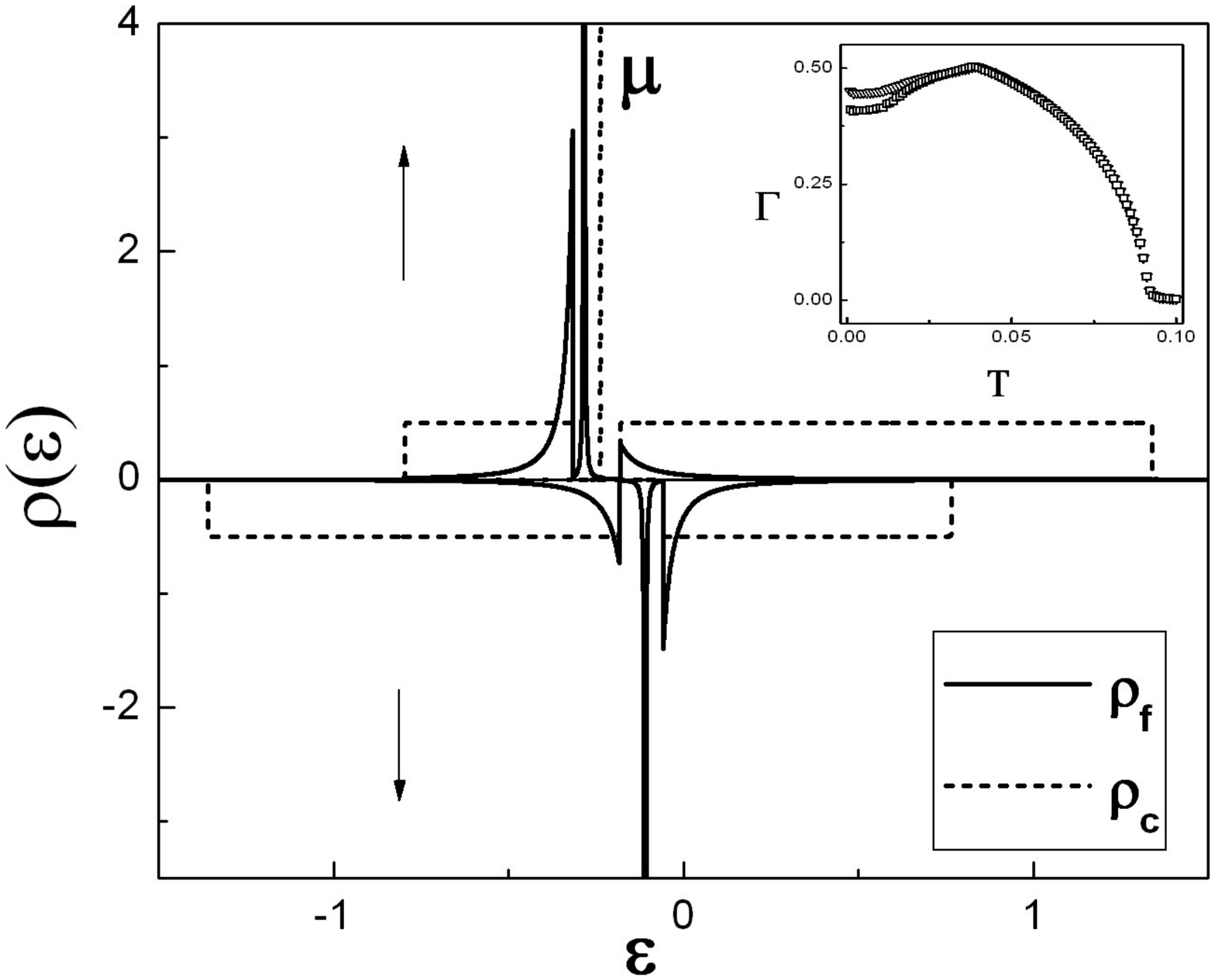}}
\caption{Schematic plot of the density of states for the magnetic
case at $T=0$. The parameters are $J_K=0.8$, $J_H=-0.01$ and
$n_c=0.8$. The continuous line corresponds to the $f$-density of
states, the dashed line to the conduction density of states and the
dotted vertical line indicates the position of the Fermi energy. The
peak of the localized f-level lies inside the gap and here the up
localized f-level is occupied and the down localized f-level is
empty. The inset shows the variation of the hybridization gap for
the two spin directions as a function of the temperature. }
\label{density0}
\end{figure}

Fig.\ref{density0} shows the $f$ and $c$ DOS  at $T=0$. The
parameters are the same as in Figure \ref{temp}. For these
parameters,  both  the magnetic order and the Kondo effect are
present in the ground state. We see from the figure that the bands
for the two spin polarizations are shifted, as expected for a
magnetically ordered state. The hybridization gap due to Kondo
effect is present for both directions of the spin, but they do not
coincide neither in the borders nor in the width. The non-hybridized
$f$-level lies inside this gap. This localized $f$-level is occupied
for spin-up states and it is empty for spin-down states. We caution
that the finite width for the two localized $f-$ levels in the
figure is due to the fact that we added a very small but finite
imaginary part to the energies. The inset of Fig.\ref{density0}
shows the temperature variations of the hybridization gaps for two
spin directions. The two gaps are nearly equal because they are
proportional to $\lambda_\sigma$, and the difference between
$\lambda_{\uparrow}$ and $\lambda_{\downarrow}$ is rather small.

\begin{figure}[t]
\centerline{\includegraphics[width=8.cm, clip=true]{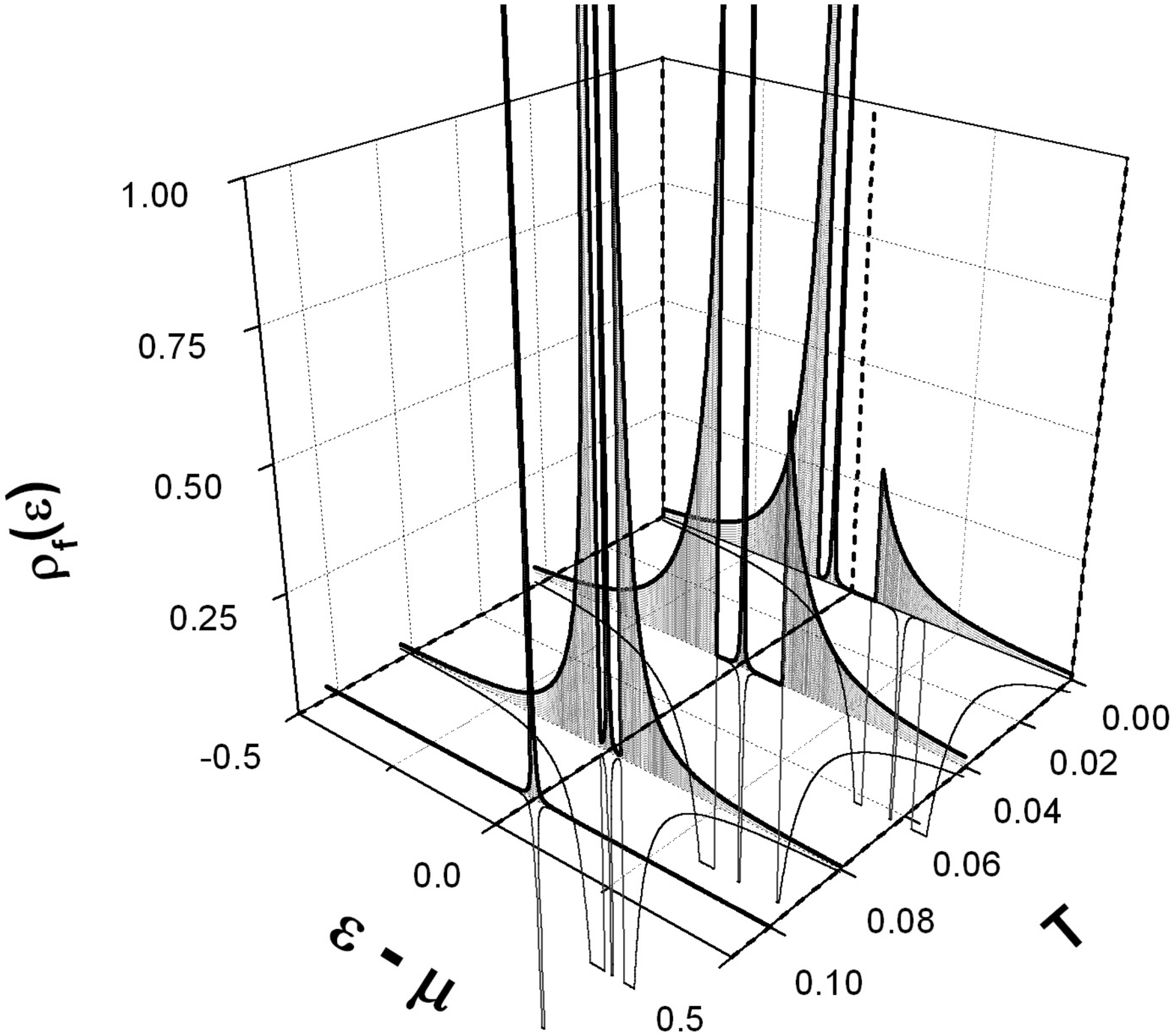}}
\caption{Plot of the up- and down- f-density of states for $T=0$ and
three finite temperatures}\label{densT}
\end{figure}

In Fig.\ref{densT} we plot the density of states at four different
temperatures: a) $T=0$ and b)a low-temperature phase ($T<T_C$) both
exhibiting coexistence of magnetic order and Kondo behavior, c) pure
Kondo phase at $T_C<T<T_K$ and d) uncorrelated high-temperature
phase at $T>T_K$. In the coexistence region at $T<T_C$, the Fermi
level remains inside the hybridization gap for the spin-up band, and
inside the conduction band, $E^{\downarrow}_{-}(k)$, for the
spin-down band. That implies a semi-metallic behavior. When $T>T_C$,
the up- and down-spin bands coincide, the Fermi level lies inside
the gap and coincides with the energy of the $f$-level, $E_{o
\sigma}$. That implies an insulating behavior. Finally, when
$T>T_K$, $\lambda_\sigma$ vanishes and there is no more coupling
between $f$ and $c$ electrons: the hybridization gap closes and the
system becomes metallic.

\subsection{Effective mass.}

\begin{figure}[t]
\centerline{\includegraphics[width=9.cm,
clip=true]{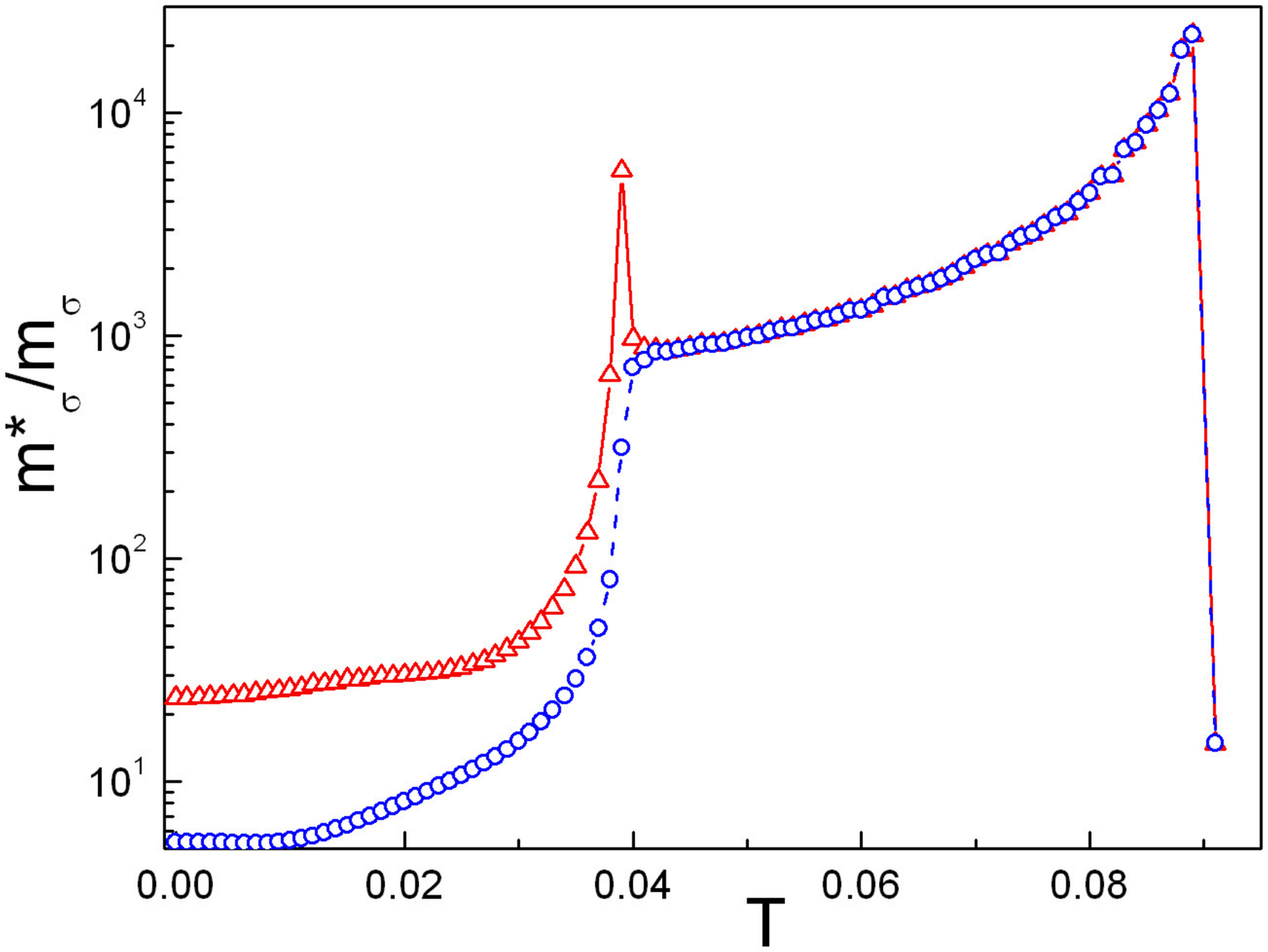}} \caption{Plot of the effective mass
versus temperature with the same parameters as in Fig.\ref{temp}.
The triangles correspond to up- spin and the circles to down- spin.}
\label{effmass}
\end{figure}

>From the quasiparticle spectrum (\ref{spectrum}) one can estimate
the mass enhancement\cite{Barzykin}:
\begin{eqnarray}
\frac{m_{\sigma}^{*}}{m}= 1 +
\frac{2\alpha^2_{\bar{\sigma}}}{(E_{o\sigma}-\mu-\Delta_{\sigma})^2}
\label{mass}
\end{eqnarray}
We plot the temperature variation of the mass enhancement in
Fig.\ref{effmass}. Two peaks are clearly seen: one corresponds to
the Curie temperature, and the second (at a higher temperature) to
the onset of Kondo effect. In the region of coexistence, the
effective mass increases as a function of temperature tracking the
position of the Fermi level inside the $E^{\bar{\sigma}}_{-}(k)$
band. In the pure Kondo phase at $T_C<T<T_K$, the mass enhancement
becomes quite large. Such a large enhancement is the consequence of
the fact that the denominator in Eq.~(\ref{mass}) goes to zero in
the pure Kondo regime, and the effective mass formally diverges. A
finite width of the $f-$level eliminates the divergence, but still
yields a very large mass. Although this estimate of the mass
probably overestimates the strength of the enhancement, it still
provides a qualitatively good explanation of the heavy fermion
behavior.

\subsection{Specific Heat}

The thermodynamic properties of the UKL model can be  easily
calculated from the expression for the free energy
Eq.~(\ref{freeen}). Here, we present only results for the specific
heat which could be compared with available experimental data. We
use the usual definition of the entropy of the system

\begin{equation}
S(T)=-\frac{\partial F}{\partial T}=-\frac{\partial }{\partial T}
(-T\ln Z)
\end{equation}
and the specific heat can be calculated as
\begin{eqnarray}
C_V(T) & = & T \frac{\partial S}{\partial T}\\ \nonumber
 & = & \frac{1}{T^2}\sum_{k, \sigma} [\frac{(E^{\sigma}_{-}({\bf
k})-\mu)^2e^{\beta (E^{\sigma}_{-}({\bf k})-\mu)}}{(1+e^{\beta
(E^{\sigma}_{-}({\bf k})-\mu)})^2}+\\
\nonumber  &  &  \frac{(E^{\sigma}_{+}({\bf k})-\mu)^2e^{\beta
(E^{\sigma}_{+}({\bf k})-\mu)}}{(1+e^{\beta (E^{\sigma}_{+}({\bf
k})-\mu)})^2}] + \\
\nonumber & & \frac{1}{T^2}\sum_{\sigma} \frac{(E_{o
\sigma}-\mu)^2e^{\beta (E_{o \sigma}-\mu)}}{(1+e^{\beta (E_{o
\sigma}-\mu)})^2}
\end{eqnarray}
\begin{figure}[t]
\centerline{\includegraphics[width=9.cm, clip=true]{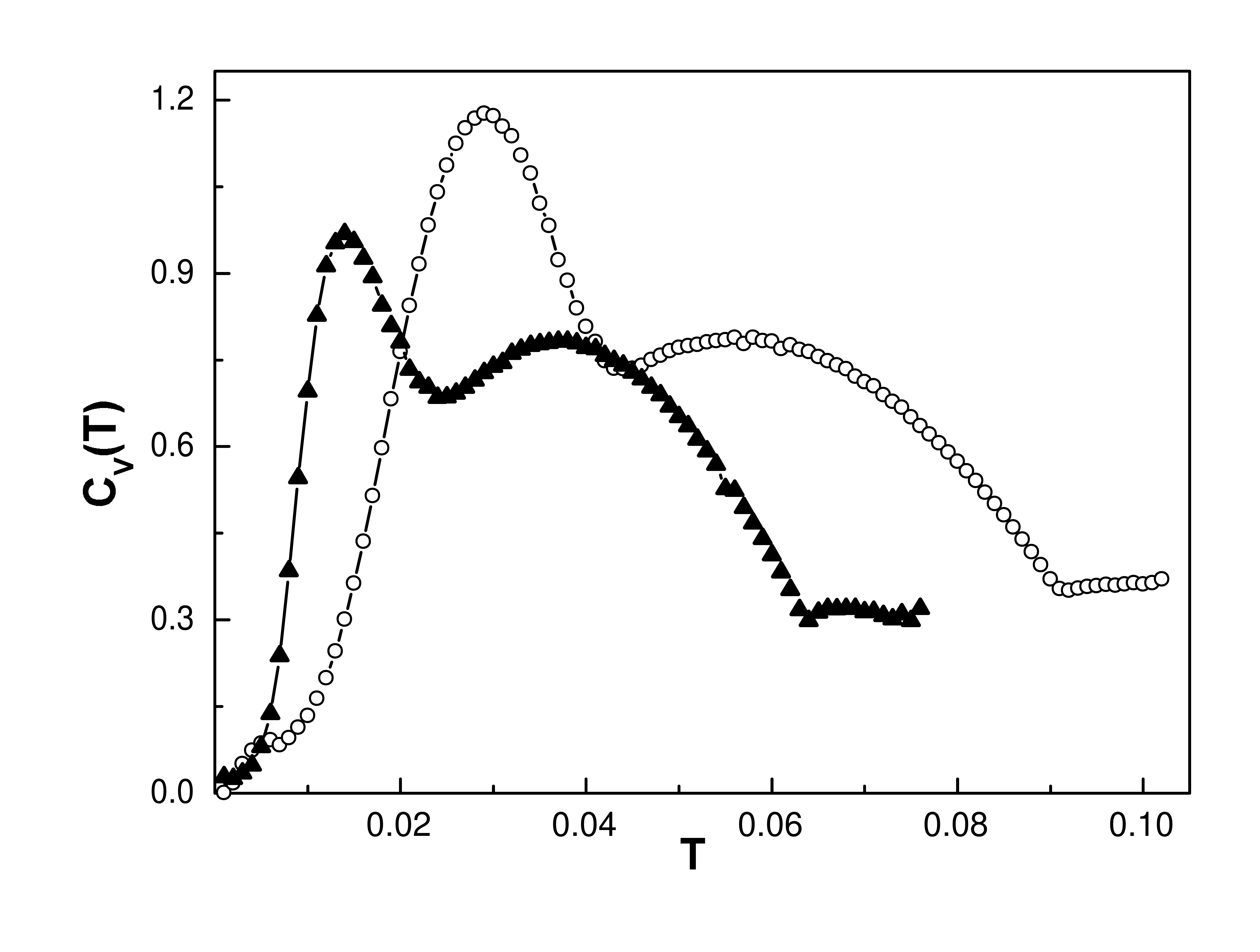}}
\caption{Plot of the electronic specific heat for the parameters a)
$J_K=0.8$, $J_H=-0.01$, $n_c=0.8$ (circles) and b) $J_K=0.7$,
$J_H=-0.001$ and $n_c=0.8$ (triangles). } \label{heat}
\end{figure}

In Fig.~\ref{heat} we present numerical results for the specific
heat as a function of temperature. As the specific heat is very
sensitive to the choice of the parameters,  we plot it for two
different sets: a) $J_K=0.8$, $J_H=-0.01$, $n_c=0.8$ (circles), and
b) for $J_K=0.7$, $J_H=-0.001$, $n_c=0.8$ (triangles). For the later
choice of parameters, the specific heat temperature dependence is
more pronounced. First, at low temperatures, the specific heat
increases with increasing $T$, following a power law with an
exponent close to $2$. Second, it goes through a peak, which can be
associated with the temperature when both $\lambda_{\uparrow}$ and
$\lambda_{\downarrow}$ become equal and to the Curie temperature. At
higher temperature, in the non-magnetic region, the specific heat is
very high because the Fermi level coincides with the localized
$f$-level, as it was discussed for the effective mass. Those values
are so high that they mask a second peak, expected for the Kondo
temperature. Therefore, when the gap in the density of states closes
at the Kondo temperature, only a change in the slope is observed. In
the case a) both peaks are "rounded up" because of the higher values
of the order parameters compared to the case b).

\subsection{Ferromagnetic Doniach diagram}

\begin{figure}[t]
\centerline{\includegraphics[width=9.cm, clip=true]{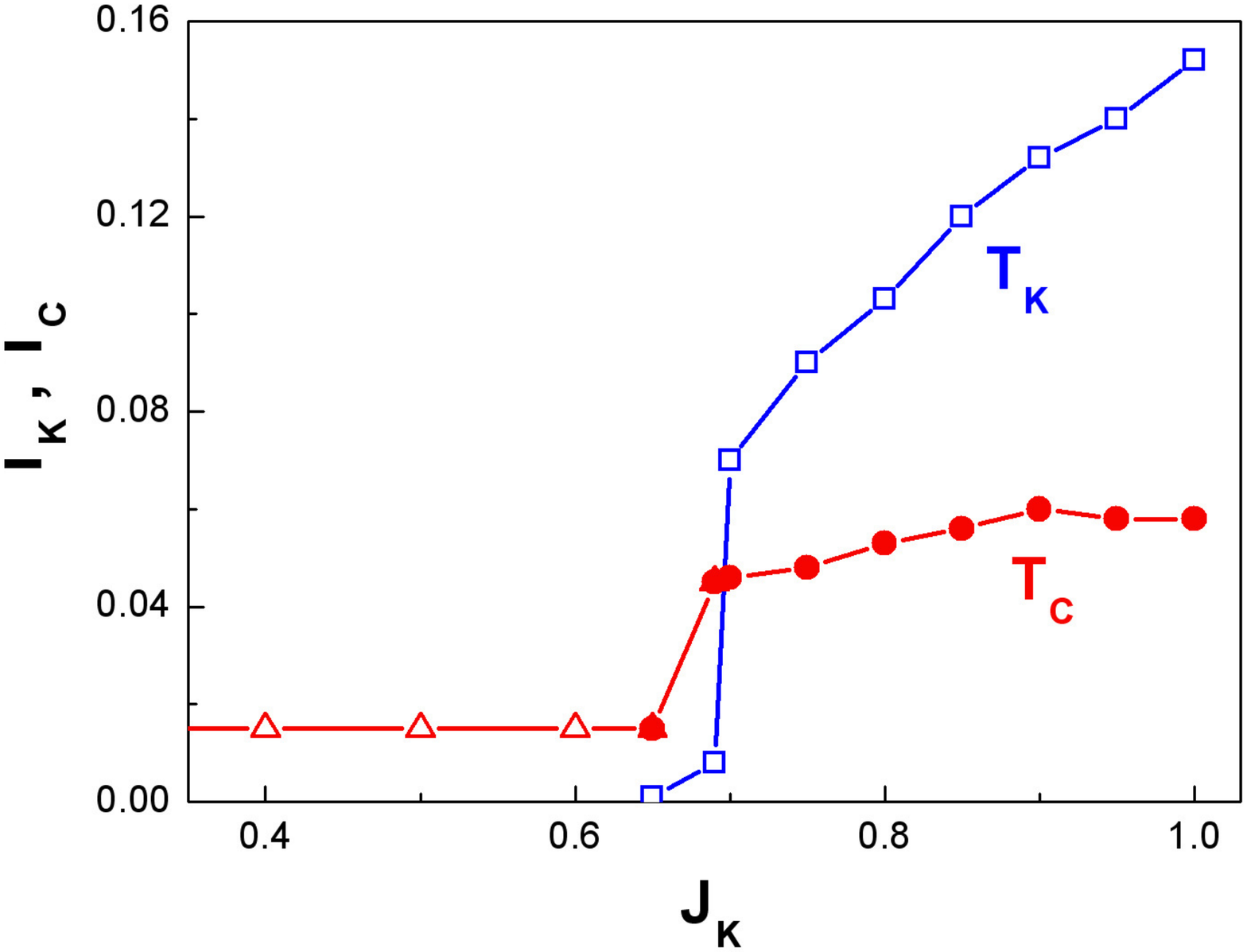}}
\caption{The ferromagnetic Doniach diagram : Plot of the Curie
temperature $T_C$ (the triangles are analytical values, the circles
numerical ones) and the Kondo temperature $T_K$ (squares) versus
$J_K$ with $J_H=-0.01$ and $n_c=0.8$.} \label{Doniach}
\end{figure}

We present here the ferromagnetic ``Doniach diagram'' for the UKL
model. In Fig.~\ref{Doniach}, we plot  the Curie temperature, $T_C$,
and the Kondo temperature, $T_K$, versus $J_K$ for fixed values of
$J_H$ and $n_c$. It is possible to see that the Kondo temperature
$T_K$ becomes finite only at the critical value $J_{K}^{c}\thicksim
0.65$ for $J_H=-0.01$ and $n_c=0.8$, then rapidly increases for
larger values of $J_K$. On the other hand, the Curie temperature,
$T_C$, is finite for all studied values of $J_K$. The two curves
$T_K(J_K)$ and $T_C(J_K)$ cross slightly above $J_{K}^{c}$ and for
larger values of $J_K$ the Kondo temperature, $T_K$, is always
larger than $T_C$. Indeed, the ferromagnetic order persists for all
values of the ratio $J_K/J_H$, while the Kondo-ferromagnetism
coexistence exists only for sufficiently large values of this ratio.

In the purely magnetic region the Curie temperature can be easily
evaluated, and it is equal to $T_C=\frac{z |J_H|}{4}$. Also, the
Kondo temperature exhibits an almost linear behavior as a function
of $J_K$. This is in contrast with the Kondo impurity case, where an
exponential dependence on $J_K$ is observed~\cite{lacroix}, and also
with the $S=1/2$ KL model where both exponential and non-exponential
behaviors can be obtained~\cite{CoqblinKLM}.

It is worth to note that the present results indicate that the Curie
temperature increases as a function of $J_K$, and this is also in
the opposite direction of the standard Doniach's diagram. This can
be understood because in the Doniach's diagram (localized spins
S=1/2), the reduction of the N\'eel temperature at large $J_K$ is
the consequence of the competition between magnetic order and Kondo
effect. The ordering temperature goes to zero at high enough values
of $J_K$. Here, there is no such competition, and the Kondo
interaction actually favors a ferromagnetic ordering through the
RKKY interaction (even if the localized magnetic moments are
partially screened by the conduction electrons). That explains the
augmentation of the Curie temperature. Then, if in the standard KL
model the short-range antiferromagnetic correlations between
neighboring $f$-moments reinforce the Kondo effect\cite{coleman,
Iglesias}, in the UKL model it is the Kondo effect that reinforces
the ferromagnetic interaction.

The diagram presented in Fig.~\ref{Doniach} can be called the
``ferromagnetic Doniach diagram'' for the UKL model and it is
qualitatively very different from the well-known Doniach diagram
derived for the Kondo lattice  with $S=1/2$, where the ordering
temperature tends to zero at a finite $J_K$ and the magnetic order
and the Kondo effect compete rather than co-exist.

\section{Conclusions.}

In conclusion, in this work we introduced and studied the
underscreened Kondo lattice model with localized spins $S=1$. In the
framework of this model, we analyzed the coexistence of
ferromagnetism and Kondo behavior. To this end, we derived and
solved a set of self-consistent equations for the relevant bosonic
fields. We obtained a region in the space of parameters, where the
order parameters, $\lambda_\sigma$, $M$, and $m$ are simultaneously
different from zero. This coexistence persists up to finite
temperatures, then the magnetic order disappears first, at the
temperature $T_C$. At this temperature $\lambda_\sigma$ is still
different from zero. In fact it is at its maximum value, indicating
that the Kondo effect is the strongest. This is, in fact, in
agreement with experiments for uranium compounds in which the Kondo
behavior is observed above the Curie temperature. The phase
transition at $T_C$ is, therefore, a transition from the
ferromagnetic Kondo phase at low $T$ to the non-magnetic Kondo phase
at higher $T$. Another transition can be obtained at $T=0$ by
varying the Kondo coupling parameter $J_K$. This transition is
discontinuous and leads to a non-Kondo magnetically ordered state
(see Fig.~\ref{lambda1}).

We emphasize that the coexistence of ferromagnetism and Kondo effect
is due to the presence of two $f-$levels: one non-hybridized and one
hybridized levels (see Fig.~\ref{density0}). The presence of the
non-hybridized $f-$level is at the origin of the partial screening
of the localized spins. Indeed, this is the key difference between
our UKL model and $S=1/2$ Kondo lattice model. In the latter, there
is only one $f-$level and it is always hybridized with the
conduction band in the Kondo regime. The result is a competition
between magnetism and Kondo effect which is reflected in Doniach
diagram.

Our UKL model can explain the behavior of some uranium compounds
such as the previously described $UTe$, $UCu_{0.9}Sb_{2}$ and
$UCo_{0.5}Sb_{2}$ compounds, which order ferromagnetically at a
large Curie temperature, $T_C$, and present a Kondo behavior. Such a
Kondo-ferromagnetism coexistence has been recently observed in other
uranium compounds. We would like to mention $UNiSi_2$ with a Curie
temperature $T_C= 95 K$~\cite{A1, A2, A3}, $UCo_{0.6}Ni_{0.4}Si_2$
with $T_C = 62 K$~\cite{B1} and $URu_{2-x}Re_{x}Si_{2}$ compounds
where $T_C$ increases rapidly with concentration
$x$~\cite{Dalicha,Torik,Bauer}. In these compounds $T_C$ increases
rapidly with $x$ and there are clear evidences of the coexistence
between the ferromagnetic order and a Non-Fermi-Liquid
behaviour\cite{Bauer}. A more extended analysis of the UKL model
can, possibly, describe also this phenomenum. Our UKL model can
finally be applied to the case of the recently observed neptunium
compound $NpNiSi_{2}$, which becomes ferromagnetic at $T_C = 51.5 K$
and presents Kondo behaviour~\cite{C1}. In the neptunium based
compounds, $5f$ electrons are relatively well localized and the
magnetic moment of Np can be described by a localized spin larger
than $S=1/2$, again corresponding to underscreened case.

In conclusion, the UKL is a very good tool to describe some
ferromagnetic-Kondo compounds and can be considered as an
improvement with respect to the regular Kondo lattice model to
account for some actinide $5f$ compounds.

\section{Acknowledgements}

NBP thanks MPIPKS, Dresden, Germany, where this work has been
partially performed and the hospitality and support of IF-UFRGS and
FAPERGS, Porto Alegre, Brazil. BC thanks the European COST P16 for
financial support. JRI acknowledges the hospitality of MPIPKS,
Dresden, Germany and LPS, UPS, Orsay, France and financial support
from Brazilian agencies CNPq, CAPES and FAPERGS.



\section{Appendix}

In this Appendix we present the derivation of the Green functions
(Eqs. (\ref{gf11})) using the equations of motion (Eqs.(\ref{gf1})).

We first derive the Green functions for the f-electrons:
\begin{eqnarray}
\begin{array}{l}
(\omega-E_{0\sigma})F_{\alpha\alpha}^{\sigma}({\bf k})=
1+\alpha_{\bar{\sigma}} \frac{\alpha_{\bar{\sigma}}
(F_{\alpha\alpha}^{\sigma}({\bf k})+F_{\beta\alpha}^{\sigma}({\bf
k}))} {\omega-\varepsilon_{{\bf k}\sigma} }
\\[0.2cm]
(\omega-E_{0\sigma})F_{\beta\alpha}^{\sigma}({\bf k})=
\alpha_{\bar{\sigma}}
\frac{\alpha_{\bar{\sigma}}(F_{\alpha\alpha}^{\sigma}({\bf
k})+F_{\alpha\alpha}^{\sigma}({\bf k}))} {\omega-\varepsilon_{ {\bf
k}\sigma} }
\end{array}
\label{gf2}
\end{eqnarray}
\noindent where we call $A_{\bar{\sigma}}=
\frac{\alpha^2_{\bar{\sigma}}} {\omega-\varepsilon_{{\bf  k}\sigma}
}$. Then, Eq. (\ref{gf2}) becomes
\begin{eqnarray}
\begin{array}{l}
(\omega-E_{0\sigma}-A_{\bar{\sigma}})F_{\alpha\alpha}^{\sigma}({\bf
k})= 1+A_{\bar{\sigma}}F_{\beta\alpha}^{\sigma}({\bf k})
\\[0.2cm]
(\omega-E_{0\sigma})F_{\beta\alpha}^{\sigma}({\bf k})=
A_{\bar{\sigma}}(F_{\alpha\alpha}^{\sigma}({\bf
k})+F_{\beta\alpha}^{\sigma}({\bf k}))
\end{array}
\label{gf3}
\end{eqnarray}
Solving this set of equations, we obtain
\begin{eqnarray}
\begin{array}{l}
F_{\alpha\alpha}^{\sigma}({\bf k})=
\frac{\omega-E_{0\sigma}-A_{\bar{\sigma}}}
{(\omega-E_{0\sigma})(\omega-E_{0\sigma}-2A_{\bar{\sigma}})}
\\[0.2cm]
F_{\beta\alpha}^{\sigma}({\bf k})=F_{\alpha\alpha}^{\sigma}({\bf
k})-\frac{1}{\omega-E_{0\sigma}}
\end{array}
\label{gf4}
\end{eqnarray}
The poles of $F_{\alpha\alpha}^{\sigma}({\bf k})$ determine the
spectrum $E^{\sigma}_{\pm}({\bf k})$, given by eqs. (\ref{spectrum})
in the main text. In terms of $E^{\sigma}_{\pm}({\bf k})$, the Green
functions of $f-$electrons are:
\begin{eqnarray}
\begin{array}{l}
F_{\alpha\alpha}^{\sigma}({\bf k})=
\frac{\omega-E_{0\sigma}-A_{\bar{\sigma}}}
{(\omega-E_{0\sigma})(\omega-E_{0\sigma}-2A_{\bar{\sigma}})}
=\\[0.2cm]
\frac{1} {2(\omega-E_{0\sigma})}-\frac{1} {2W_{\sigma}({\bf k})} [
\frac{\varepsilon_{{\bf k}\sigma}-E^{\sigma}_{+}({\bf k})}
{\omega-E^{\sigma}_{+}({\bf k})}- \frac{\varepsilon_{{\bf
k}\sigma}-E^{\sigma}_{-}({\bf k})} {\omega-E^{\sigma}_{-}({\bf k})}]
\end{array}
\label{gf5}
\end{eqnarray}
\noindent and
\begin{eqnarray}
\begin{array}{l}
F_{\beta\alpha}^{\sigma}({\bf k})=F_{\alpha\alpha}^{\sigma}({\bf
k})-\frac{1}
{\omega-E_{0\sigma}}=\\[0.2cm]
-\frac{1} {2(\omega-E_{0\sigma})}-\frac{1} {2W_{\sigma}({\bf k})} [
\frac{\varepsilon_{{\bf k}\sigma}-E^{\sigma}_{+}({\bf k})}
{\omega-E^{\sigma}_{+}({\bf k})}- \frac{\varepsilon_{{\bf
k}\sigma}-E^{\sigma}_{-}({\bf k})} {\omega-E^{\sigma}_{-}({\bf
k})}]~,
\end{array}
\label{gf6}
\end{eqnarray}
\noindent where $ W_{\sigma}({\bf k})\equiv E^{\sigma}_{+}({\bf
k})-E^{\sigma}_{-}({\bf k})$.

We next calculate $G_{c\alpha}^{\sigma}({\bf k}) $. Using the
corresponding equation of motion
\begin{eqnarray}
\begin{array}{l}
(\omega-\varepsilon_{{\bf  k}\sigma})G_{c\alpha}^{\sigma}({\bf k})=
\alpha_{\bar\sigma}(F_{\alpha\alpha}^{\sigma}({\bf
k})+F_{\beta\alpha}^{\sigma}({\bf k}))~,
\end{array}
\label{gf7}
\end{eqnarray}
\noindent we obtain:
\begin{eqnarray}
\begin{array}{l}
G_{c\alpha}^{\sigma}({\bf k})=
\frac{\alpha_{\bar\sigma}}{(\omega-\varepsilon_{{\bf k}\sigma})
(\omega-E_{0\sigma}-2A_{\bar{\sigma}})}=
\\[0.2cm]
\frac{\alpha_{\bar\sigma}} {W_{\sigma}({\bf k})} [ \frac{1}
{\omega-E^{\sigma}_{+}({\bf k})}- \frac{1}
{\omega-E^{\sigma}_{-}({\bf k})}]
\end{array}
\label{gf8}
\end{eqnarray}

Finally, the Green functions for itinerant electrons
$G_{cc}^{\sigma}({\bf k}) $ are calculated in the same way, starting
with:
\begin{eqnarray}
\begin{array}{l}
(\omega-\varepsilon_{{\bf k}\sigma }) G_{cc}^{\sigma}({\bf k})=
\alpha_{\bar\sigma}(G_{\alpha c}^{\sigma}({\bf k})+G_{\beta
c}^{\sigma}({\bf k}))+1
\\[0.2cm]
(\omega-E_{0\sigma})G_{\alpha c}^{\sigma}({\bf k})=
\alpha_{\bar\sigma}G_{cc}^{\sigma}({\bf k})
\end{array}
\label{gf9}
\end{eqnarray}
And solving eqs. (\ref{gf9}), we obtain
\begin{eqnarray}
\begin{array}{l}
G_{cc}^{\sigma}({\bf k})= -\frac{1} {W_{\sigma}({\bf k})} [
\frac{E_{0\sigma}-E^{\sigma}_{+}({\bf k})}
{\omega-E^{\sigma}_{+}({\bf k})}-
\frac{E_{0\sigma}-E^{\sigma}_{-}({\bf k})}
{\omega-E^{\sigma}_{-}({\bf k})}]
\end{array}
\label{gf10}
\end{eqnarray}


\begin{thebibliography}{99}
\bibitem{f} A ferromagnetic coupling
between localized magnetic moments may result from two
contributions: an effective RKKY interaction and, possibly, a direct
exchange. The RKKY interaction may lead to ferromagnetic ordering
for strong coupling regime, large $|J_K|$, and only for partial
filling of the conduction  band, as it has been shown by many
authors for the usual KL model, see, for example,
refs.~\onlinecite{Tsurevmod,honner}.
\end{thebibliography}
\end{document}